\documentclass[manuscript]{acmart}

\AtBeginDocument{
  \providecommand\BibTeX{{
    \normalfont B\kern-0.5em{\scshape i\kern-0.25em b}\kern-0.8em\TeX}}}

\usepackage{amsmath}
\usepackage{hyperref}
\usepackage{booktabs}
\usepackage{comment}
\usepackage{longtable}

\usepackage{url}
\usepackage{algorithmic}
\usepackage{graphicx}
\usepackage{textcomp}
\usepackage{xcolor, soul} 
\usepackage{tabularx}
\usepackage{enumitem}
\usepackage{tcolorbox}
\usepackage[english]{babel}
\usepackage{hyphenat}
\usepackage{float}
\usepackage{multirow}

\usepackage{array}
\usepackage{caption}

\newboolean{showcomments}
\setboolean{showcomments}{true}
\ifthenelse{\boolean{showcomments}}
{\newcommand{\nb}[2]{
		\fcolorbox{black}{yellow}{\bfseries\sffamily\scriptsize#1}
		{\sf\small$\blacktriangleright$\textit{#2}$\blacktriangleleft$}
	}
	
}
{\newcommand{\nb}[2]{}
	
}

\begin{document}

\title{Architectural Anti-Patterns in Student-Developed Microservice Architectures: An Exploratory Study}

\author{Anna Rita Fasolino}
\author{Marco De Luca}
\author{Michele Perlotto}
\author{Porfirio Tramontana}
\affiliation{%
  \institution{University of Naples Federico II}
  \department{Department of Computer Engineering}
  \city{Naples}
  \country{Italy}
}

\email{fasolino@unina.it}
\email{marco.deluca@unina.it}
\email{michele.perlotto@unina.it}
\email{ptrmont@unina.it}

\begin{abstract}
Teaching microservice software architectures is challenging due to distributed design complexity and the gap between classroom settings and industrial practice. Comprehending the quality issues that students tend to introduce when implementing microservice architectures (MSAs) is crucial for improving educational strategies. 

This work investigates the quality of student-developed microservice architectures using a reference taxonomy of MSA anti-patterns and derives a set of lessons learned along with actionable teaching recommendations.

We conducted a longitudinal, project-based educational experience across three editions of a Software Architecture Design course (2023–2025) in which 216 Master’s students (67 teams) collaboratively designed, evolved, and deployed a realistic, containerized, web-based MSA for a gamified software-testing platform. The final feature-complete system was analyzed to identify recurring design errors, investigate their causes, and extract insights for improving future course iterations.

A total of 23 anti-patterns out of the 58  in the taxonomy were detected, meaning that nearly half of the recognized MSA design issues appeared in the students' projects. 
These anti-patterns spanned five categories. \textit{Security} problems were the most common, revealing recurring challenges in implementing authentication, authorization, and data protection mechanisms. \textit{Team Organization} issues followed, reflecting students’ limited experience with DevOps practices and collaborative workflows. \textit{Service Interaction} problems reflected challenges in inter-service communication and coordination. Fewer issues emerged in \textit{Intra-service Design} and \textit{Inter-service Decomposition} categories, suggesting that students were generally more successful in establishing service boundaries and structuring services internally. 

Overall, our findings offer an empirical snapshot of quality issues in student-built MSAs and provide guidance for instructors aiming to help students design more robust, maintainable, and secure systems. A consistent observation was that students tended to prioritize feature delivery over resilience, robustness, and operational discipline. To counter this tendency, we recommend adopting minimal, enforced standards such as API contracts and gateway/discovery conventions, providing hands-on labs on asynchronous and resilient communication, incorporating security-by-design practices (e.g., least privilege, secret management, encrypted service-to-service traffic), and offering ready-to-use CI/CD templates. 

Finally, the paper presents a comprehensive and realistic educational experience in designing and implementing a full-fledged microservice-based architecture, a contribution not yet represented in the existing literature. By outlining both the technical outcomes and the pedagogical insights, it offers a replicable model that may inspire instructors and educators seeking to incorporate authentic, industry-aligned MSA projects into their courses.

\end{abstract}




\maketitle

\section{Introduction}

In recent years, awareness of the importance of teaching software architectures has grown considerably \cite{Lago2005}. 
Software architecture can be defined as the fundamental organization of a software system, including its components, their interactions, and the guiding design principles \cite{ISO42010}. Understanding software architecture is essential, as it enables students to reason about system structure, make informed design decisions, and ensure qualities such as scalability, maintainability, and evolvability in complex systems.

However, there are many difficulties encountered both by teachers and by students when dealing with software architecture education \cite{Galster2016}. These include the abstract nature of architectural concepts, the lack of opportunities for hands-on experience with sufficiently complex systems, the challenge of objectively achieving and assessing architectural quality, and the gap between academic exercises and real-world industrial practices.
One of the most important and critical aspects to teach is how to develop quality software architectures, that is, how to translate required quality attributes into concrete design decisions. Well-established frameworks for instructors are available to address this topic, such as the one presented in the book by Bass, Clements, and Kazman \cite{Bass2003}, which provides not only architectural patterns and styles, but also a detailed set of design tactics capable of ensuring the achievement of specific architectural quality attributes. 
Nevertheless, achieving high-quality software architecture is not solely a matter of applying styles, patterns, or design practices. While these approaches are essential, they are not sufficient, as developers can still introduce flawed solutions, commonly referred to as code smells or anti-patterns \cite{fowler1999refactoring}, which can compromise the quality of the resulting system. 

Recently, teaching microservice architectures (MSAs) has attracted growing interest within the scientific community, reflecting their widespread adoption and success in industry \cite{Ferreira2025TeachingMS, TeachingMSADevOps2022}. Microservice architectures \cite{fowler2014microservices} are increasingly used to develop a wide range of applications, such as web platforms, enterprise systems, and cloud-native solutions, and it is becoming increasingly important to ensure their quality \cite{Schirgi2021}.
Several architectural and code smells can generally affect even microservice-based architectures, often resulting from improper service decomposition, excessive coupling, suboptimal service communication, or violations of domain boundaries \cite{8354414}. Making students aware of these issues is therefore crucial to help them develop better design decisions and cultivate good software engineering practices in MSA-based projects. A way to better convey software architecture design and quality concepts to students is project-based learning \cite{GUO2020}, a well-known learning approach requiring that students have the opportunity to apply learned principles in real projects \cite{Fioravanti2018}. 
Without practical experience, indeed, students risk developing only a superficial understanding of the issues at hand. 
However, providing realistic software development experiences is challenging for various reasons, not least the limited time available to design and implement a non-trivial software architecture within the duration of a single-term course.

Probably due to the difficulties mentioned above, only a few small-scale experiences with project-based learning in the field of microservice architectures have been reported in the literature \cite{Ferreira2025TeachingMS, TeachingMSADevOps2022, Qian2025OBEMicroservices}, and there is still insufficient research investigating quality issues in the architectures produced by students. In particular, little is known about the types of design mistakes, architectural smells, and anti-patterns that students tend to introduce when implementing microservice architectures, and how these issues relate to common quality attributes such as maintainability, scalability, and modifiability.
Understanding these quality problems is crucial for improving educational strategies, as it allows instructors to provide targeted guidance, highlight common pitfalls, and design interventions that reinforce good design principles. Moreover, analyzing the patterns of errors in student projects can inform the development of teaching materials, exercises, and assessment methods that emphasize not only theoretical knowledge but also practical, high-quality software design.
Therefore, investigating the recurring quality issues in student-developed microservice architectures is essential for bridging the gap between theoretical instruction and real-world application, ensuring that students acquire both conceptual understanding and practical skills necessary to produce maintainable and robust systems.

To address this gap, we conducted a longitudinal, project-based learning experience on software architecture design, involving 67 student teams over three consecutive editions of a Software Architecture Design course within a Computer Engineering Master’s program. Unlike typical short-term student projects, which are often small, simplified, and unrealistic, this approach allowed teams to develop, evolve, integrate, test, and refactor a realistic microservice-based application over time, closely simulating industrial software development.
At the conclusion of the projects, we assessed the quality of the resulting architecture against established microservice code smells and anti-patterns, identifying common quality issues, analyzing their causes, and deriving lessons learned to inform and improve future teaching practices. This study provides insights into the typical architectural challenges faced by students and highlights strategies for improving education in microservice-based software development.

Its main contributions are:
\begin{itemize}
   \item We design and conduct a longitudinal, project-based learning experience on microservice architecture design with 216 students across three course editions, using a realistic microservice-based application and providing a replicable model for instructors seeking to integrate industry-aligned MSA projects into their courses.
    \item We empirically analyze the resulting system to identify recurring architectural anti-patterns in student-developed microservice architectures, relating them to common quality attributes such as maintainability, scalability, and modifiability.
    \item We investigate the underlying causes of these quality issues, highlighting the role of both architectural decisions and process/DevOps practices in shaping the resulting architecture.
    \item We distill a set of lessons learned and actionable recommendations for teaching microservice architectures, aimed at helping instructors better support students in designing high-quality, microservice-based systems.
    
\end{itemize}

The remainder of the paper is organized as follows:
Section \ref{sec:Background} provides background on microservice architectures, common code smells and architectural anti-patterns, and techniques for their detection and analysis. Section \ref{sec:education} describes the project-based teaching experience in software architecture design. Section \ref{sec:Study} details the study conducted to identify anti-patterns and quality issues introduced by students in the developed microservice projects. Section \ref{sec:Results} presents the results of the quality analysis, including classifications of detected code smells and anti-patterns. Section \ref{sec:discussion} interprets the findings and discusses their implications for MS-based architecture education. Section \ref{sec:Related} reviews related work and Section \ref{sec:Conclusions} concludes the paper and outlines directions for future work.

\section{Background}
\label{sec:Background}

This section outlines the main concepts underlying our study. It introduces microservice architectures and their core principles, distinguishes between code smells and architectural anti-patterns, and reviews existing techniques and tools for detecting such issues in microservice-based systems.

\subsection{Microservices Architectures}

Originally conceived as an evolution of monolithic architectures, which are traditionally characterized by tight coupling, limited modularity, and difficulties in achieving scalability, the microservice architectural style has gained significant traction since around 2011 \cite{fowler2014microservices, newman2015building}. Its adoption has been driven by the need for greater flexibility, scalability, and maintainability in modern software systems, particularly in contexts requiring frequent updates, continuous delivery, and integration with cloud-native environments.

Key principles and characteristics of microservice architectures include:
\begin{itemize}
    
\item Service independence: each microservice encapsulates a specific bounded context, enabling modularity, independent development, testing, deployment, and scaling. This separation supports teams working autonomously and reduces the risk of changes propagating across the system.

\item Decentralization: microservices are autonomous, often employing different technologies, programming languages, and databases that best suit their individual requirements \cite{dragoni2017microservices}. This technological heterogeneity allows for experimentation and optimization within each service but requires careful design of communication patterns and integration strategies.

\item Fault isolation and resilience: failures are contained within individual services, preventing cascading effects across the system. Techniques such as circuit breakers, retries, and bulkheads are commonly employed to enhance system robustness.

\item Automation through CI/CD and containerization: continuous integration, continuous delivery, and container-based deployment pipelines allow rapid, reliable, and repeatable software evolution \cite{soldani2018pains}. This automation reduces human error and accelerates release cycles.

\item Well-defined APIs: interactions between services occur through standardized, versioned APIs, ensuring interoperability, backward compatibility, and clear separation of concerns. This promotes integration across services and supports long-term maintainability.

\end{itemize}
The main advantages of MSA include enhanced scalability, maintainability, and resilience, accelerated delivery cycles, and technological freedom, allowing teams to adopt the best tools and frameworks for each service. However, these benefits come with notable challenges. The distributed nature of microservices introduces complexity in communication, latency, and data consistency, requiring careful management of distributed transactions and eventual consistency. Additional challenges include communication overhead, operational maturity, difficulties in service governance, monitoring, and observability \cite{niedermaier2019observability}, and the coordinated evolution of service contracts and APIs.

Furthermore, while MSA is widely adopted in industry, there is growing interest in understanding how its principles and challenges translate into educational settings. Teaching students to design, develop, and maintain microservice-based applications requires not only conveying technical knowledge but also helping them navigate distributed complexity, service coordination, and resilient system design, which are often underestimated in small-scale, short-term educational projects. Examining student-developed microservices provides insight into common architectural pitfalls, bridging the gap between theoretical knowledge and practical skills necessary for professional software development.

\subsection{Code Smells and Architectural Anti-Patterns in MSA}
The adoption of microservice architectures has introduced new design challenges, leading to the identification of specific code smells and architectural anti-patterns. These issues often result from violations of core microservice principles, such as loose coupling, bounded contexts, and domain-driven design.

Code smells are generally defined as symptoms in the source code or design that may indicate deeper problems \cite{fowler1999refactoring}. They do not necessarily represent defects, but rather design issues that can hinder maintainability, scalability, or evolvability if left unresolved.

In microservices, common examples of code smells include:
\begin{itemize}
    
\item God Service: services that grow too large and encompass multiple responsibilities, violating the single responsibility principle.

\item Improper communication patterns: over-reliance on synchronous calls, chatty APIs, or inconsistent messaging protocols.

\end{itemize}

In contrast, anti-patterns are recurring poor solutions to common design problems \cite{brown1998antipatterns}, which explicitly capture flawed practices that negatively affect system quality.
In MSAs, anti-patterns often emerge at a higher architectural level, such as:
\begin{itemize}
    
\item Shared Persistence: multiple services accessing the same database schema, creating hidden coupling and transactional issues.

\item Cyclic Service Dependencies: services that call each other in a loop, making deployment and evolution complex.

\item Hardcoded Configuration or Secrets: reducing security and flexibility.

\end{itemize}

However, the distinction between code smells and anti-patterns is not always clear-cut. Several studies report and classify both code smells and anti-patterns in MSAs \cite{Taibi2019MicroservicesAntiPatterns}, \cite{Tighilt2020}, but these classifications frequently overlap, and the terminology is sometimes used inconsistently or ambiguously \cite{Meissner2021EAsLiT}. For example, God Service may be categorized either as a service-level code smell or as an architectural anti-pattern depending on the perspective adopted. This ambiguity highlights the importance of carefully interpreting existing taxonomies when analyzing the quality of microservice-based systems \cite{8354414}.

One reference point in this context is the classification of microservice anti-patterns presented in \cite{Cerny2023MicroserviceAntiPatterns}. This work, based on a tertiary study of the literature, systematically organizes the smells and anti-patterns identified in prior literature reviews into a single comprehensive catalog. The catalog includes 58 disjoint anti-patterns derived from 203 originally reported issues. 
The anti-patterns are grouped into five categories,
described in Table \ref{tab:placeholder}, which also reports the total number of anti-patterns for each category.

\begin{itemize}
    
\item \textit{Intra-service design}: issues within a single service, such as god classes, tight coupling of modules, or insufficient error handling.

\item \textit{Inter-service decomposition}: improper splitting of services, leading to overly coarse or overly fine-grained services, shared databases, or misaligned domain boundaries.

\item \textit{Service interaction}: poor communication patterns between services, including excessive synchronous calls, chatty APIs, or fragile orchestration.

\item \textit{Security}: weaknesses in authentication, authorization, data handling, and configuration management.

\item \textit{Team organization}: mismatches between team structure and service boundaries, lack of coordination, and poor development practices that propagate architectural problems.

\end{itemize}

\begin{table}[h!]
    \caption{Microservice anti-pattern taxonomy from \cite{Cerny2023MicroserviceAntiPatterns}}
    \centering
    \begin{tabular}{|c|p{8.5cm}|c|}
    \hline
    \textbf{Anti-pattern Category} &\textbf{ Description} & \textbf{Items} \\ \hline \hline
    Intra-service Design   & Concerns flaws in the internal structure or design of individual services (such as Nano-service, Mega-service, or Ambiguous service). & 9\\ \hline
    Inter-services Decomposition   & Concerns flaws in the structural division of the system across at least two microservices (such as Cyclic-dependency, Chatty service, or Wrong cuts). & 14\\ \hline
    Service Interaction & Concerns flaws in the communication patterns between services, focusing on violations that lead to improper communication paths, reduced system scalability, or diminished resilience (such as No API-Gateway or Hardcoded endpoints). & 9\\ \hline
    Security & Concerns flaws in authentication, authorization, and data encryption security principles (such as Unauthenticated Traffic or Publicly Accessible Microservices). &  10 \\  \hline
    Team Organization & Concerns flaws arising from development team decisions, including implementation strategies, migration planning, operations, and monitoring practices (such as Too many standards, No CI/CD, or Insufficient Monitoring). & 16 \\\hline
    \end{tabular}
    \label{tab:placeholder}
\end{table}

This classification provides a reference to micro-service developers to design better-quality systems and to researchers who aim to detect system quality based on anti-patterns. Moreover, it can be used by educators to structure curricula, provide best practices, or design teaching materials on MSA quality based on it.

\subsection{Techniques and tools for MSAs analysis and anti-pattern detection}

Several approaches for detecting anti-patterns in MSAs have been proposed in the literature.
As discussed by Bushong et al. \cite{Bushong21}, techniques for the architectural analysis of microservices can be broadly classified into different categories, each focusing on specific perspectives and data sources of the system under study.

The most established approach is \textit{static analysis}, which inspects source code without execution to identify structural flaws through metrics like complexity, coupling, and cohesion \cite{ALFAYEZ2023107147, BALDASSARRE2020106377}. Recently, Schneider et al. \cite{Cerny_tool} conducted the first comprehensive empirical comparison of static analysis tools for architecture recovery in microservice applications. \textit{Pattern-based analysis} builds upon static inspection and focuses on recognizing recurring design structures or specific rule violations that correspond to known anti-patterns, documented in established taxonomies \cite{Pigazzini2020, Marquez19, Walker20}.
\textit{Dynamic analysis} instead observes the system during execution, and by monitoring performance, latency, and component interactions, it helps detect runtime issues such as synchronization errors, bottlenecks, or cascading failures \cite{Offline_Mining, maruf2022telemetry}. 
\textit{Model-based analysis}  uses abstract models representing the microservice architecture, such as dependency matrices or stochastic representations (e.g., Continuous-Time Markov Chains), for reasoning about its properties, including reliability, availability, and maintainability \cite{Mendonca20, McZara20}. Another category of approaches is \textit{graph-based analysis}, which represents the system architecture as a graph to study dependencies, communication flows, and service relationships. Depending on how the graph is constructed, this approach can rely on static information (e.g., source code, configuration files) or dynamic traces (e.g., runtime logs, telemetry) \cite{BRANDON20, liu19}. 

A further dimension is provided by \textit{analysis based on the behavior of the development team}, which focuses on team activity and organization. Metrics such as code ownership, change frequency, and author dispersion reveal coordination issues and social complexity that may underlie technical problems \cite{CodeOwnershipPrinciples2024, CodeOwnershipAISecurity2023, AlignmentOwnershipContribution2019, SourcegraphVision2023}. 

As these works show, effective anti-pattern detection requires a hybrid multi-level strategy that integrates static, temporal, behavioral, dynamic, and organizational analyses. Such complementarity provides a more complete understanding of software quality and supports evidence-based detection of anti-patterns. Cerny et al.  \cite{Cerny2023MicroserviceAntiPatterns} address the challenges of anti-pattern detection by proposing a three-phase framework that clearly distinguishes: (1) \textit{Information extraction} phase, where static, dynamic, hybrid or manual approaches can be used to gather information of different types from the analyzed system; (2) \textit{Intermediate Representation}, where the collected information is used to derive more abstract views about the system,  and (3) \textit{Detection}, where the intermediate representations can be traversed to detect anti-patterns, possibly from a reference taxonomy.

All the contributions reported in this Background section highlight the significant attention that the literature has devoted to the quality of microservice architectures, their code smells and anti-patterns, and the challenges related to their detection.
However, understanding code smells and anti-patterns is essential not only in industrial contexts but also in educational settings, as they provide concrete guidance for identifying common pitfalls in student-developed microservice projects. By analyzing these issues in educational environments, researchers and instructors can better teach design principles, architectural thinking, and best practices, bridging the gap between theory and practice.

\section{The Software Architecture Design teaching experience}
\label{sec:education}
In this section, we present the Software Architecture Design teaching experience we conducted. We first outline the structure of the course, then describe the requirements of the software project assigned to the students, and finally detail how the project-based learning experience was implemented.

\subsection{Course organization}
Our teaching experience spanned three editions of the Software Architecture Design course taught by one of the authors at the first year of the Master's Degree in Computer Engineering at the University of Naples Federico II, 
from 2023 to 2025. The course is offered as an elective course and is worth 6 ECTS credits, consisting of 48 hours of instruction over 12 weeks, 40\% of which are devoted to practical sessions.
All students enrolled in the course have prior knowledge of object-oriented programming and software engineering principles, gained in previous Bachelor’s courses.

The general goal of this course is to give students an understanding of what software architecture is and how it is created, documented, and used in practice. The course presents fundamental concepts from a vast body of theoretical knowledge available about software architectures, including architectural styles, patterns, and tactics that help to realize quality attributes \cite{Bass2003, Taylor2009}. It also aims to present how to document an architecture using multiple views, adopting both informal notations and UML \cite{Clements2002}.
The course is complemented by agile practices, such as iterative development, Scrum-based task management, and Continuous Integration, as well as the development of soft skills, which are essential for successfully working with software architectures in industrial projects of any scale \cite{Ambler2012}. The textbooks adopted in the course are reported in Table \ref{tab:book}.

To achieve its learning objectives, the course adopts a project-based approach in which students, organized into different teams consisting of up to five persons, collaborate on the design and development of a software architecture. 
Table \ref{tab:course_schedule} summarizes the 12-week course schedule for the software architecture course. It shows the weekly topics and the type of session (lecture, lab, or project review), illustrating the balance between theoretical instruction and hands-on experience. The project development phase begins in the second half of the course, starting in Week 7, and continues uninterrupted until the end of the course. 
\begin{table}[ht]
\caption{Textbooks adopted in the course.}
\centering
\small
\begin{tabular}{p{0.35\textwidth}p{0.45\textwidth}p{0.12\textwidth}}
\hline
\textbf{Authors} & \textbf{Title} & \textbf{Reference} \\
\hline

L. Bass, P. Clements, R. Kazman &
\textit{Software Architecture in Practice} &
\cite{Bass2003} \\

R. N. Taylor, N. Medvidovic, E. M. Dashofy &
\textit{Software Architecture: Foundations, Theory, and Practice.} &
\cite{Taylor2010} \\

P. Clements, D. Garlan, L. Bass, et al. &
\textit{Documenting Software Architectures: Views and Beyond} &
\cite{Clements2002} \\

C. Richardson &
\textit{Microservices Patterns: With examples in Java} &
\cite{Richardson2018} \\
S. W.~Ambler, M. Lines &
\textit{Disciplined Agile Delivery: A Practitioner's Guide to Agile Software Delivery in the Enterprise} &
\cite{Ambler2012} \\
\hline
\end{tabular}

\label{tab:book}
\end{table}

As to the project, across all three editions of the course, it focused on the same case study, and to further motivate the students, it was realistic, requiring the development and evolution of a fully functional application

conceived within the framework of an initiative funded by a transnational education project\footnote{Anonymized Project}.

Within this project, the teams were assigned different tasks covering various aspects of software development.

The student teams were supported by two tutors, Ph.D. students in Computer Engineering, who provided both methodological and technological guidance on demand.
In line with Agile practices, student project reviews were conducted on a biweekly basis.

This setup not only provided students with a meaningful hands-on experience in collaborative software development but also resulted in a realistic and evolving project, suitable for investigating architectural quality issues and recurring design problems.

\begin{table}[h!]
\centering
\caption{Overview of the 12-week Software Architecture Design course schedule.}
\begin{tabular}{|c|p{10cm}|l|}
\hline
\textbf{Week} & \textbf{Topic} & \textbf{Type} \\ 
\hline
\hline
1  & Introduction to Software Architectures and its different Structures& Lecture \\ 
\hline
2 & Attribute Driven Design (ADD) and Architectural Quality Tactics & Lecture\\ 
\hline
3  & Introduction to Git and Software Development Workflows & Lecture + Lab \\ 
\hline
4  & Three-Tiered, MVC, and Sense-Compute-Control - Architectural Patterns & Lecture \\ 
\hline
5  & Monolithic, Layered, Repository, Data-Flow, Event-Driven  Architectural Styles & Lecture \\ 
\hline
6  &  Service Oriented and Microservice Architecture & Lecture \\ 
\hline
7 & Project Presentation and Tasks Assignment & Project Start\\ 
\hline
8  & Designing and modeling microservice architectures & Lecture + Lab \\ 
\hline
9  & API design, service communication, and integration patterns + First Project Review& Lecture + Lab \\ 
\hline
10 & Agile Practices \& Soft Skills& Lecture + Lab \\ 
\hline
11  & Containerization and deployment with Docker and CI/CD pipelines + Second Project Review& Lab \\ 
\hline
12 & Project presentations and final evaluation & Final Exams \\ 
\hline
\end{tabular}

\label{tab:course_schedule}
\end{table}

\subsection{The assigned project}
\label{sec:project}
In this section, we report the requirements of the software application to be developed by the students as their course project. 

The software application was intended to 
operationalize a gamified learning process in which students author JUnit tests for a target Java class and compete against preconfigured ``robot'' opponents, consisting of automatic generators of unit tests (e.g., Randoop\footnote{Randoop, \url{https://randoop.github.io/randoop/}} and Evosuite\footnote{Evosuite, \url{https://www.evosuite.org/}}). Points, achievements, and match history foster deliberate practice and sustained engagement. This educational setting imposes strict requirements on responsiveness, artefact traceability, and flexible curation of classes and opponents. 

As to the functionality, the system delivers four cohesive capability areas that enable the gamified software testing experience. 
First, an \emph{administration} area enables instructors to curate the catalog of testable classes and to provide bundles of ``robot'' opponents. Second, the \emph{gameplay} area offers both practice and timed single–match modes. Students iterate through compile–measure cycles with immediate feedback, compete against a selected opponent, and retain a persistent match history. Third, the \emph{measurement} area integrates coverage and testing metrics into the loop: JaCoCo coverage is produced at compilation time, while EvoSuite metrics are collected at submission. 
Finally, the \emph{operations} area provides secure user registration and authentication, scheduled backups, and end-to-end artefact traceability to support course management and grading.

The design of the architecture was guided by a set of architectural constraints derived from both educational goals and practical considerations. 
First, the system was conceived according to the \emph{microservice architectural style}. This choice was motivated by the didactic objective of exposing students to the principles of service decomposition, service autonomy, and API-based integration, which are core competencies in contemporary software engineering. Adopting microservices also enabled the organization of student teams around individual services, thereby allowing parallel development activities. Synchronization across teams was then managed through code reviews and integration sessions on a shared codebase, effectively reproducing collaborative practices observed in industrial settings.

Second, for user interaction, the system adopts the \emph{Model–View–Controller (MVC) pattern}. This pattern was selected not only for its didactic value, given that MVC is widely taught as a fundamental architectural paradigm, but also for its suitability in this context. MVC provides a clear separation between presentation, domain logic, and input control, which simplifies the management of rapidly evolving user interfaces. Moreover, the modularity offered by MVC reduces the cognitive load for students working on different layers of the application while ensuring maintainability and extensibility of the web front-end.

Third, the system was required to be entirely \emph{web-based}. This ensures accessibility without installation overhead and allows the platform to be used in heterogeneous laboratory environments and from remote locations, which is crucial for an educational tool meant to be adopted across multiple courses and institutions.

Fourth, the deployment of the architecture had to be \emph{container-based}. Containerization supports the decomposition of the application into multiple independently deployable services, lowers the friction of assembling heterogeneous technology stacks, and enhances portability across different deployment environments. These properties are essential both from a didactic perspective, since container technology was itself part of the course content, and from a practical one, as they simplify setup, scaling, and reproducibility of the platform in laboratory and classroom scenarios.

Finally, for educational purposes, and consistent with the microservices principle of technological heterogeneity \cite{fowler2014microservices,newman2021building}, we deliberately avoided imposing restrictions on the technologies adopted for the implementation of individual services. This choice not only reflects industrial practice, where polyglot architectures are common \cite{newman2021building}, but also enhances student motivation and engagement by fostering autonomy, critical thinking, and hands-on decision making, in line with established guidelines for project-based learning in software engineering education \cite{souza19, perez20, Fioravanti2018}. 

\subsection{The Project-based learning activity}

Now we describe how the project-based learning activity was conducted across the three course editions, highlighting the assignment of tasks, iterative integration of components, collaborative design reviews, and the progressive development of a fully functioning system. In total, 216 students contributed to this project, organized into 67 teams.
  
In the \textit{first course edition}, at the start of the project, the instructors clarified the required specifications and constraints, providing an initial architectural draft that outlined the styles and patterns to be used. Through a series of plenary design meetings, the student teams collaboratively proposed the core components of the architecture, defining their responsibilities and the services they were expected to provide to the other components. The instructor, acting primarily as a supportive facilitator, monitored the discussions and intervened only when strictly necessary to ensure coherence, feasibility, and alignment with the project constraints. This light-touch supervision enabled students to take ownership of the architectural decision-making process while still establishing a viable and consistent foundation for the subsequent development phases.

Each team then selected the architectural components they intended to work on, which were formally assigned to them by the instructor as their exam tasks. This initial step emphasized both autonomy and responsibility, giving students ownership of their components from the very beginning. 

In total, 9 different tasks were defined and assigned to 32 teams, with multiple teams working on the same task. This choice to intentionally repeat certain tasks was driven by organizational considerations. On the one hand, it prevented an excessive proliferation of micro-tasks; on the other, it ensured that a team's failure to deliver or complete a critical architectural component would not jeopardize the entire project.

Thanks to the microservice-based architecture, all teams were able to work in parallel on their tasks, while regular review sessions and stand-up meetings enabled them to synchronize and coordinate design decisions. These collaborative interactions encouraged reflection on trade-offs, problem-solving strategies, and the rationale behind architectural choices. 
Upon completing their tasks, students submitted their code and accompanying documentation on GitHub, along with a demo, which was subsequently evaluated for the course assessment. Feedback from tutors provided guidance for improvements and reinforced iterative learning, allowing students to refine both technical and teamwork skills. 

At the end of the course edition, since multiple versions of the same core components were available, the instructor, acting as the Software Architect of the project, supported by the tutors, decided which versions to integrate in order to develop the first consolidated version of the system.
For the selection of components to be integrated into the final course-wide version of the architecture, only minimal criteria were applied. Specifically, a component was considered suitable if it satisfied the required functional specifications, provided the necessary APIs for the intended operations, and covered the core functionalities needed by the application. Components that did not meet these essential requirements were discarded, while those that fulfilled them were integrated into the consolidated system, regardless of other quality aspects such as code style or performance.

The tutors were responsible for integrating the selected components into a final, course-wide consolidated version  \textit{V1} of the architecture. This version then served as the starting point for the teams in the following course edition.

In the \textit{second course edition}, 16 student teams first studied the assigned version of the system, using and testing it to become familiar with its structure and functionality. 
New use cases, as well as architectural modifications aimed at improving the quality attributes of the system, were collaboratively defined, supported by regular review sessions that facilitated discussion and feedback. Implementing advanced administration features, adding new patterns such as an API Gateway to the architecture, and implementing a training game mode are examples of the new features the students decided to add.
Subsequently, 6 new tasks were defined and assigned to the various teams, with replications. Upon completing their tasks, each team integrated their assigned component into the project baseline. At the end of all the submissions, the tutors were responsible for selecting and integrating the most promising components into a final, course-wide consolidated version \textit{V2} of the architecture. For the selection of components to be integrated, they used the same minimal criteria used in the first course edition.

In the \textit{latter course edition}, the same approach adopted in the second edition was followed, building upon the integrated architecture produced at the end of the previous iteration. In this edition, the teams mainly introduced additional features, including new game modes, new gamification elements such as awarding prizes and achievements to players, and social functionalities that allowed players to share their game experiences within a network of friends. These new requirements were assigned to the 19 student teams as 6 evolutionary tasks, following the same process of development, review, and integration established in the earlier editions.

Throughout the experience, students were actively engaged in planning, implementing, testing, and coordinating, fostering not only technical proficiency in microservice architecture but also essential soft skills such as teamwork, communication, and project management. By the end of the course, the iterative, collaborative process enabled the students to deliver and deploy a fully functioning web application that successfully implemented the core functionalities initially required. This tangible outcome highlighted both the practical learning achieved and the students’ ability to navigate complex, real-world software development challenges.

Table \ref{tab:project} summarizes the number of teams, students, and tasks per course edition.
\begin{table}[ht]
\centering
\caption{Project Teams, Students, and Tasks per Course Edition.}
\begin{tabular}{lrrrr}
\toprule
\textbf{Course Edition} & \textbf{Team\#}  & \textbf{Student\#} & \textbf{Task\#} \\
\midrule
First          & 32 & 116 & 9 \\
Second          & 16  & 46  & 6\\
Third   & 19  & 54   & 6    \\
\midrule
\textbf{Total} & \textbf{67} & \textbf{216} & \textbf{21} \\
\bottomrule
\end{tabular}
\label{tab:project}
\end{table}

\section{Assessing the Architectural Quality of the Microservice-based System}
\label{sec:Study}

Thanks to the educational experience we carried out, we were able to setup our study on the architectural quality of student-developed software architectures. The specific study goal was to identify the occurrence of anti-patterns in microservice-based applications developed by students. 

To classify MSA anti-patterns, we decided to use the taxonomy proposed by Cerny et al. \cite{Cerny2023MicroserviceAntiPatterns} that is described in  Table \ref{tab:placeholder}.

Guided by this categorization, we decided to investigate the following Research Questions:

\begin{enumerate}[label=RQ\arabic*, ref=RQ\arabic*]
\item Which ``\textit{Intra-service Design}" anti-patterns occur in microservice architectures developed by students?

\item Which ``\textit{Inter-service Decomposition}" anti-patterns occur in microservice architectures developed by students?

\item Which ``\textit{Service Interaction}" anti-patterns occur in microservice architectures developed by students?

\item Which ``\textit{Security}" anti-patterns occur in microservice architectures developed by students?

\item Which ``\textit{Team organization}" anti-patterns occur in microservice architectures developed by students?

\end{enumerate}

Together, these Research Questions allow us to obtain a comprehensive picture of the types and frequencies of architectural issues that emerge when novice developers design and implement microservice-based systems.

To answer these research questions, we designed our study following the structured framework proposed by Wohlin \textit{et al.} \cite{wohlin2024experimentation}. According to this framework, we defined the objects and subjects involved in the experiment, outlined the experimental procedure, and specified the data analysis and validation steps undertaken to ensure the reliability of our results.

\subsection{Experimental Objects}

The experimental objects represent the entities being studied or manipulated during the experiment. We selected the final third-year release of the student-developed web application as our experimental object because it represents the most complete and stable version of the system. Focusing on a single, fully integrated and functioning implementation ensured a consistent basis for identifying architectural issues, code smells, and anti-patterns. In contrast, analyzing multiple partial or incomplete versions would have introduced variability due to missing features or non-executable components, making it harder to distinguish genuine design flaws from artifacts of incomplete development. Moreover, many architectural and interaction-related anti-patterns can only be reliably observed in a functioning, integrated microservice system.

The selected version implemented the main gamification features, including two different game modes, leaderboard displays, player profile management, and administrative functions intended for the teacher. 

Its architecture consisted of seven microservices deployed on Docker containers, all connected to a common Docker network. The system also included two gateways, likewise containerized and connected to the same network, one responsible for serving static assets and handling client-side routing (i.e., UIGateway), and the other managing backend-level routing (i.e., API Gateway).

\begin{figure}[h!]
  \centering
  \includegraphics[width=0.9\linewidth]{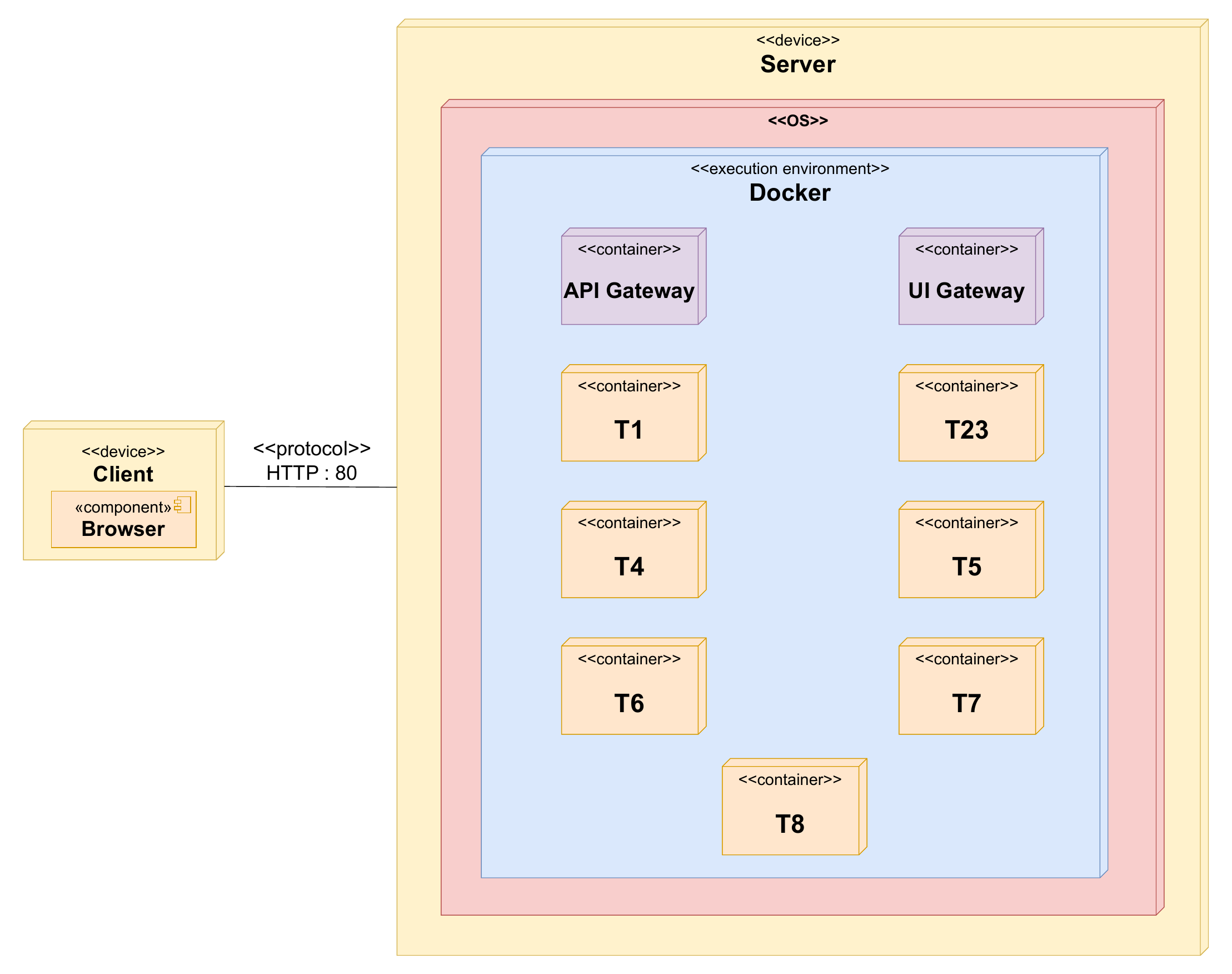}
  \caption{Deployment Diagram of the Web Application.}
  \label{fig:deploy}
\end{figure}

Figure~\ref{fig:deploy} illustrates the deployment view of the architecture and the seven containerized microservices (i.e., \texttt{T1}, \texttt{T23}, \texttt{T4}, \texttt{T5}, \texttt{T6}, \texttt{T7}, \texttt{T8}), whereas Table~\ref{tab:components} refines this view by summarizing each component’s main responsibilities and the technologies used for its implementation. Complementing these structural details, Table~\ref{tab:webapp} provides a concise characterization of the web application in terms of programming languages and size metrics, reporting for each language the number of files and non-commented lines of code (NCLOCs), which together offer a compact view of the implementation complexity. 

\begin{table}[h!]
\centering
\caption{Architectural Components Responsibilities and Implementation Technologies.}
\begin{tabular}{p{4cm} p{3cm} p{5cm}}
\toprule
\textbf{Component} & \textbf{Technology} & \textbf{Brief responsibility} \\
\midrule
UI Gateway & NGINX & Static asset provision; reverse proxy; initial URL-based routing to API gateway.\\ \hline
API Gateway & Spring Cloud Zuul & Central routing, REST API authentication.\\ \hline
T1: Admin Service & Spring Boot MVC (Thymeleaf), Spring Security & Curating testable classes and robot bundles; listing admins/players; providing class artefacts to gameplay and related UIs. \\ \hline
T23: Student Service & Spring Boot MVC, Spring Security (JWT/BCrypt) & Student Registration/Authentication/ Credentials management, student profile management, and related UI.\\ \hline
T4: Game Repository & Go & Persistent data about player matches, timestamps, and gameplay outcome data.\\ \hline
T5: Game Front End & Spring Boot MVC & Player Home UI; Game Presentation UI. \\ \hline
T6: Game Engine & Spring Boot REST & Handling game logic. \\ \hline
T7: Student Test Runner & Spring Boot REST & Unit Test compilation, execution, and JaCoCo-based coverage evaluation. \\ \hline
T8: EvoSuite Runner  & Node.js & EvoSuite test generation; EvoSuite-based player's test metrics evaluation.\\ \hline
\bottomrule
\end{tabular}

\label{tab:components}
\end{table}

\begin{table}[ht]
\centering
\caption{Implementation footprint of the web application, showing file counts and NCLOC per language.}
\begin{tabular}{lrrr}
\toprule
\textbf{Language} & \textbf{Files} & \textbf{NCLOC} \\
\midrule
CSS         & 21  & 12663  \\
HTML         & 33  & 4375  \\
Java         & 65  & 5635  \\
JavaScript   & 25  & 2995  \\
XML         & 8 & 589  \\
YAML         & 4 & 58  \\
Go         & 28  & 3683  \\
Dockerfile  & 10  & 120       \\
Shell   &  8  & 325 \\
\midrule
\textbf{Total} & \textbf{202} & \textbf{30443} \\
\bottomrule
\end{tabular}
\label{tab:webapp}
\end{table}

The project is of medium size, comprising 202 files and approximately 30,000 effective lines of code. It primarily uses \texttt{Go} and \texttt{Java} for application logic, with front-end components in \texttt{HTML}, \texttt{CSS}, and \texttt{JavaScript}. The repository also contains configuration artifacts such as Dockerfiles and \texttt{YAML} scripts, as well as auxiliary \texttt{Shell} scripts.\footnote{Due to the necessity of anonymizing the submission, at the moment we cannot provide the link to the GitHub project repository, but we intend to share it in case of paper publication.}.

\subsection{Subjects}

The subjects of this study were the 216 students who contributed to the development of the web application across the three editions of the Software Architecture Design course.
As reported in Section \ref{sec:project}, all the students were enrolled in the first year of a two-year Master’s program in Computer Engineering. Since most of them held a Bachelor’s degree from the same university, they shared a similar academic background. They had already completed one Software Engineering course during their previous studies, as well as several programming courses covering languages such as C, C++, Java, and Python. However, to better characterize the student cohort, we collected demographic information and previous experiences with specific software technologies, frameworks, and tools to be used in the project. To this aim, we used an anonymous survey administered at the start of each course edition. 

According to the data collected by the survey, the average age of participants was 23 years. Among respondents, 85\% self-identified as male and 15\% as female. As to their previous experiences, 
Table \ref{tab:survey_results} reports the data collected by the survey. 
\begin{table}[ht]
\centering
\small
\caption{Summary of students’ prior experience and background (N = 216).}
\begin{tabular}{llrr}
\toprule
\textbf{Question} & \textbf{Option} & \textbf{n} & \textbf{\%} \\
\midrule

\multicolumn{4}{l}{\textbf{Web development experience (HTML/CSS/JS frameworks)}} \\
\midrule
 & None                                 & 58  & 27 \\
 & Small exercises / personal projects  & 108  & 50 \\
 & $\geq$1 or more university projects  & 28  & 13 \\
 & Extra-university / work experience   & 22  & 10  \\
\midrule

\multicolumn{4}{l}{\textbf{Spring MVC / Spring Boot experience}} \\
\midrule
 & Never                                & 199 & 92 \\
 & Only small exercises                 & 4  & 2 \\
 & $\geq$1 university project           & 7  & 3 \\
 & Extra-university / work experience   & 6   & 3  \\
\midrule

\multicolumn{4}{l}{\textbf{Rest API experience}} \\
\midrule
 & Never                                & 43 & 20 \\
 & Theory only                          & 128  & 59 \\
 & $\geq$1 or more university projects  & 32  & 15 \\
 & Extra-university / work experience   & 13   & 6  \\
\midrule

\multicolumn{4}{l}{\textbf{Middleware programming experience}} \\
\midrule
 & None                                 & 59 & 27 \\
 & Theory only                          & 86  & 40 \\
 & $\geq$1 or more university projects  & 67  & 31 \\
 & Extra-university / work experience   & 4   & 2  \\
\midrule

\multicolumn{4}{l}{\textbf{Docker / containerization experience}} \\
\midrule
 & Never                                & 136 & 63 \\
 & Exercises / lab only                 & 56  & 26 \\
 & $\geq$1 full project                 & 15  & 7  \\
 & Extra-university / work experience   & 9   & 4  \\
\midrule

\multicolumn{4}{l}{\textbf{Work experience in software development}} \\
\midrule
 & None                                 & 201 & 93 \\
 & $\geq$1 internship / job             & 15  & 7 \\
\bottomrule
\end{tabular}
\label{tab:survey_results}
\end{table}
 
Students had limited exposure to Web technologies: 27\% reported no previous experience, while 50\% had only completed small exercises in a prior programming course. Only a minority (8\%) reported some familiarity with Spring-based frameworks, which are not covered in the standard Bachelor’s curriculum, whereas 92\% reported no experience at all.
With respect to REST APIs, most students (59\%) had only theoretical knowledge, primarily acquired in an optional Bachelor-level course offered at the same university. A smaller subset (21\%) reported more substantial experience with REST APIs from previous projects (often Bachelor’s theses) or professional work, whereas 20\% declared no prior knowledge at all. A similar pattern emerged for middleware: the majority of students had either purely theoretical knowledge (40\%) or none (27\%), while 31\% reported previous academic exposure, typically from an optional advanced programming course. The use of Docker was also largely new to the cohort: 63\% of students had never used it before, and only 11\% reported experiences with Docker in non-trivial projects. The limited extra-academic work experience was further corroborated by the answers to the final question, which indicated that only 7\% of students had completed an internship or a work experience in software development before taking this course.

Overall, the survey confirms that students had limited practical exposure to the technologies used in the project. Most had only basic or theoretical familiarity with Web development, REST APIs, middleware, and Docker, and very few had prior professional experience. This homogeneity in background supports the validity of interpreting the observed issues as representative of challenges typically encountered by novice developers.

\subsection{Experimental Procedure}

To analyze the software architecture and detect the anti-patterns present in the system, we adopted a three-phase process defined according to the detection framework proposed by Cerny et al. \cite{Cerny2023MicroserviceAntiPatterns}. The procedure was structured as follows:
(i) \textit{Information extraction}, in which we collected up-to-date structural and behavioral evidence from source code, repositories, and runtime artifacts;
(ii) \textit{Intermediate representation generation}, where heterogeneous inputs were transformed into analyzable models and graphs to consolidate a system-level view; and
(iii) \textit{Anti-pattern detection}, where rule-based and metric-based analyses, complemented by human-in-the-loop validation, were applied to identify and categorize microservice anti-patterns.

In each phase, we employed available static and dynamic analysis tools to support our investigation, as well as custom prototypes developed specifically for this study. In the following Section \ref{sec:Results}, we detail how each phase was conducted and report the principal findings.

\subsection{Data Analysis and Validation  }

Across the entire process, from information gathering to model construction and anti-pattern identification, all activities and intermediate artifacts were produced independently by two authors and then validated by a third, who also adjudicated disagreements. Inputs and outputs were versioned and time-stamped to ensure traceability, with validation notes and decision rationales recorded to maintain an auditable trail. Cases supported by weak or incomplete evidence were discarded to avoid false positives. This dual coding with third-author validation reduced subjectivity, improved consistency, and strengthened construct validity and reproducibility. Finally, all authors collaboratively reflected on the interpretation of the data and the presentation of the results. 

\section{Results}
\label{sec:Results}

This section presents the results obtained from the three phases of the experimental procedure.
Each subsection describes the evidence collected, the quantitative outcomes, and the main observations supporting the identification of architectural issues. %”

\subsection{Information Extraction}

In this phase, we extracted the evidence required for anti-pattern analysis. Following Cerny et al. \cite{Cerny2023MicroserviceAntiPatterns}, we prioritized up-to-date sources over static documentation. The project documentation was fragmented across student teams and produced at different times, increasing the risk of inconsistency; by contrast, source code and runtime artifacts offer authoritative views of the system’s current structure and behavior. Accordingly, we grounded our analysis in the codebases of all architectural components and complemented them with selected operational traces and targeted manual inspection. Our inputs comprised the microservices’ source repositories (application code and tests), build and deployment descriptors (e.g., container manifests, gateway and routing configuration), available API specifications, and centralized logs collected under representative execution scenarios.

The analysis proceeded in three steps. First, we ran static analysis on every component with the \textsc{SonarQube} tool\footnote{SonarQube, \href{https://www.sonarsource.com/products/sonarqube/} {https://www.sonarsource.com/products/sonarqube/}} to establish a baseline of code quality and potential technical debt,  and complemented this with the reverse-engineering and code-navigation facilities of IntelliJ~IDEA\footnote{IntelliJ IDEA, \href{https://www.jetbrains.com/idea/}{https://www.jetbrains.com/idea/}}, which supported the extraction of class relationships, internal dependency flows, and controller–service–repository interactions of microservices developed in Spring Boot MVC\footnote{Spring Boot MVC, \href{https://spring.io/guides/gs/serving-web-content}{https://spring.io/guides/gs/serving-web-content}}.

Second, to capture the system’s execution topology, we performed dynamic analysis by exercising representative user flows and collecting centralized application logs and HTTP traces, exposing them through \textsc{Grafana}\footnote{Grafana, \href{https://grafana.com/}{https://grafana.com/}} dashboards. This observability setup enabled us to reconstruct temporal call sequences and inter-service dependencies, as well as to monitor endpoint-level latency and error rates, thus identifying runtime coupling and potential bottlenecks. These runtime traces provided the evidence needed to distinguish purely structural anomalies from anti-patterns that manifest only under load or along specific interaction paths (e.g., chatty communication or emergent bottlenecks).

Finally, we performed a focused manual review to close information gaps: inspecting gateway routes, authentication filters, and service-to-service configuration; aligning inferred endpoints with available API descriptions; and validating ambiguous findings against the existing documentation. Documentation was used only to disambiguate code and trace-based inferences, not as a primary source.

This phase produced a consolidated evidence set for the next stage: (i) per-service static analysis snapshots (\textsc{SonarQube} findings, derived indicators, and the structural insights obtained through IntelliJ~IDEA’s reverse-engineering features), (ii) a catalog of endpoints and observed inter-service calls reconstructed from logs and traces, and (iii) configuration-level facts about routing, authentication, and deployment. These artifacts constitute the raw material that we transform into intermediate representations in the subsequent phase, enabling systematic anti-pattern detection at both service and system levels.

\subsection{Intermediate Representations Generation}

We used the information collected in the previous \textit{Information Extraction} phase to reconstruct several views about the analyzed system.

\textit{First}, we aggregated the \textsc{SonarQube} outputs into a component-level quality report that consolidates, for each microservice, structural complexity indicators and maintainability-related proxies. Table~\ref{tab:repo-stats} provides an overview of the implementation footprint of each component by reporting, for every programming language used within a service, the number of files, the total non-comment lines of code (NCLOC), and the percentage that each language contributes to the component’s overall codebase. The \textit{UI Gateway} component is not represented in this analysis, as it is implemented solely through an Nginx configuration file (\texttt{nginx.conf}) and does not contain executable application source code that can be meaningfully characterized in terms of NCLOC. This characterization offers a concise picture of the system’s technological heterogeneity and of the relative size of each component, helping us to identify polyglot or unusually large services that are more prone to intra-service design or decomposition anti-patterns. 

\begin{figure}[h!]
  \centering
  \includegraphics[width=0.8\linewidth]{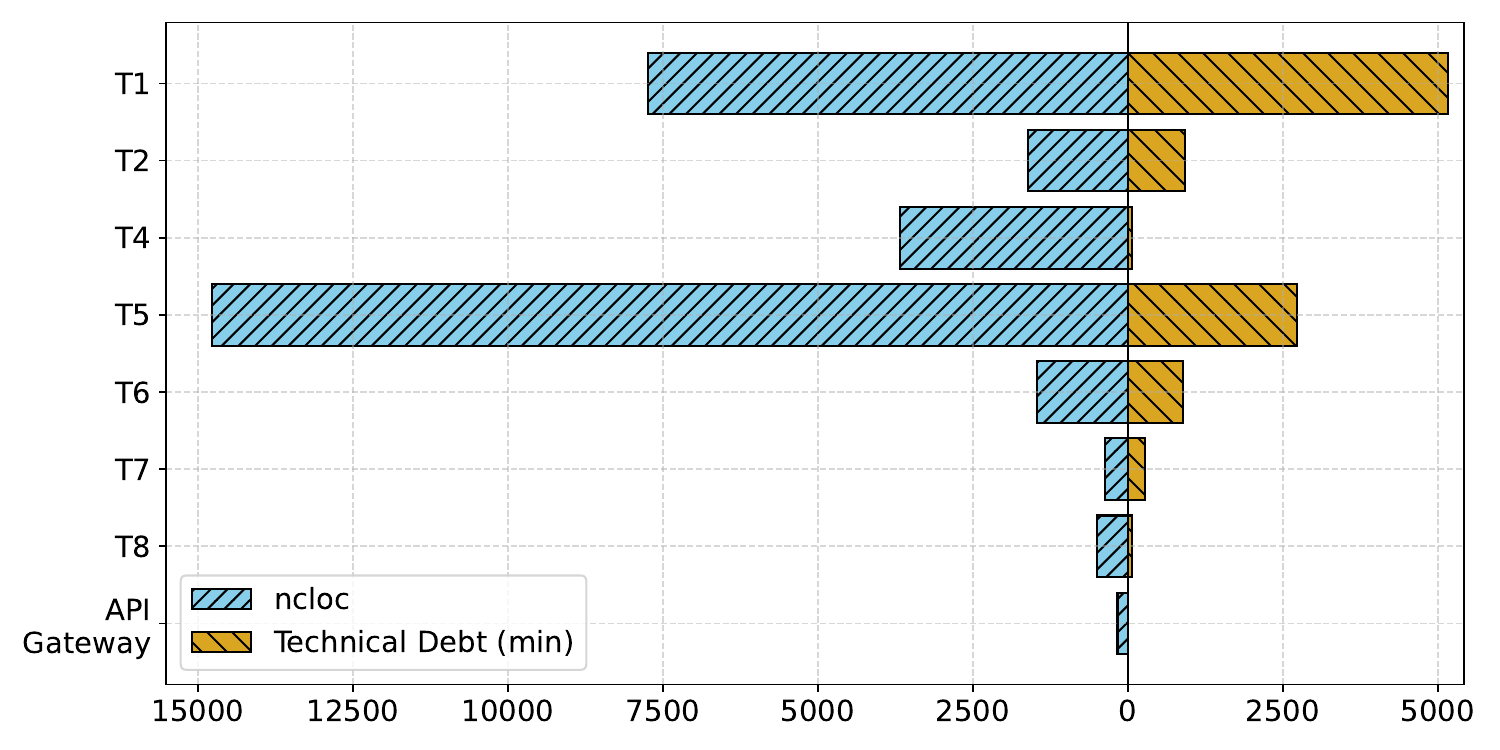}
  \caption{Modules' Technical Debt computed by SonarQube and Module Size.}
  \label{fig:sonar_td}
\end{figure}

Figure~\ref{fig:sonar_td} complements this view by juxtaposing, for each microservice, the Technical Debt (TD) estimated by SonarQube with its size in terms of NCLOC, thus distinguishing components whose TD is mainly size-driven from those where TD is high relative to their footprint. For instance, \texttt{T5} is the largest service (14{,}779~NCLOC) and exhibits substantial TD (2{,}726~min), whereas \texttt{T1} combines a smaller size (7{,}750~NCLOC) with the highest TD (5{,}166~min), indicating a higher density of issues. In contrast, \texttt{T7}, \texttt{T8}, and the \texttt{API Gateway} remain small with low TD. Together, these component-level views highlight the main maintenance hotspots and provide a first indication of which services should be prioritized for detailed inspection in the subsequent anti-pattern detection steps.

\textit{Second}, we reconstructed per-service detailed views (e.g., UML Class Diagrams, Package Diagrams, and Activity Diagrams) to expose each component’s internal structure and clarify its processing and responsibilities. These models were derived from the structural evidence gathered in the previous phase and were supported by the reverse-engineering and code-navigation features of IntelliJ~IDEA, which facilitated the extraction of class relationships and dependency flows. The resulting views make explicit the internal structure, control flow, and allocation of responsibilities, providing a basis for reasoning primarily about intra-service design and decomposition anti-patterns (e.g., over-large or ambiguous services, misplaced responsibilities), while also offering complementary evidence for security- and interaction-related issues (e.g., where authentication, data access, or cross-service coordination logic is concentrated).

\begin{figure}[h!]
  \centering
  \includegraphics[width=\linewidth]{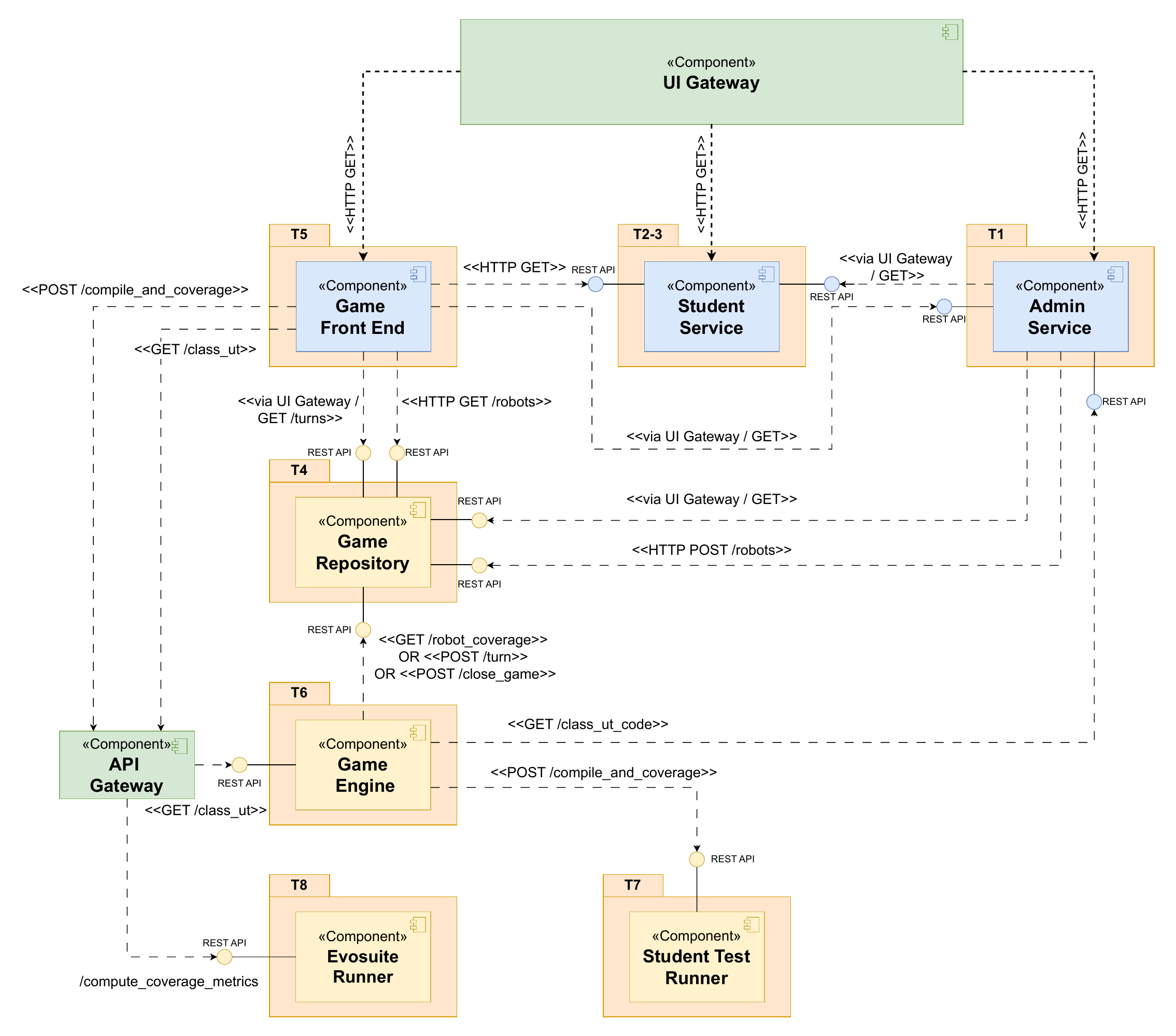}
  \caption{High-level microservice architecture.}
  \label{fig:architecture}
\end{figure}

Moreover, we created a UML Component-and-Connector (C\&C) view to provide an overview of the system’s complete architecture. It was obtained via manual reverse engineering of gateway routes, configuration files, and observed HTTP traces. The latter were inspected and correlated through \textsc{Grafana} dashboards, which helped validate the direction and frequency of inter-service calls and highlight actual runtime communication paths. This system-level model, depicted in Figure~\ref{fig:architecture}, captures interfaces, components, and inter-component dependencies, and thus supports the detection of interaction-related anti-patterns such as centralized bottlenecks or excessively chatty communication among services. From this representation, we distinguish three categories of modules: headless microservices (\texttt{T4}, \texttt{T6}, \texttt{T7}, \texttt{T8}), microservices also exposing a GUI (\texttt{T1}, \texttt{T23}, \texttt{T5}), and the two gateway modules (\texttt{UI Gateway}, \texttt{API Gateway}).

\begin{table}[ht]
\centering
\caption{Overview of analyzed component with languages and code size.}
\begin{tabular}{llrr}
\hline
\textbf{Component} & \textbf{Languages} & \textbf{NCLOC} & \textbf{Files} \\
\hline
T1 & CSS (11,5\%), HTML (31,7\%), Java (44,9\%), JS (10,5\%), Shell (0.2\%), XML (1,2\%) & 7750 & 63 \\
T23 & CSS (29\%), HTML (21\%), Java (38\%), JS (7\%), XML (6\%) & 1612 & 29 \\
T4 & Go (100\%) & 3683 & 30 \\
T5 & CSS (77\%), HTML (5\%), Java (5\%), JS (13\%), XML (1\%) & 14779 & 45 \\
T6 & HTML (65\%), Java (30\%), XML (6\%)& 1464 & 5 \\
T7 & Java (62,8\%), Shell (0,5\%), XML (36,1\%), YAML (0,5\%) & 366 & 6 \\
T8 & JS (31,3\%), XML (7,8\%), Shell (60,9\%) & 501 & 10 \\
API Gateway & Java (55\%), XML (45\%) & 168 & 4 \\
\hline
\end{tabular}

\label{tab:repo-stats}
\end{table}

Collectively, these intermediate representations provide a consistent basis for the detection rules and metrics described in the next section.

\subsection{Anti-pattern Detection}

Several automated techniques for anti-pattern detection in microservice architectures have been proposed in the literature \cite{maruf2022telemetry, Mendonca20, Pigazzini2020, BRANDON20}. 
However, for known limitations \cite{Cerny2023MicroserviceAntiPatterns}, automated analysis does not suffice for detecting any anti-pattern.
Consequently, in this phase, we had to complement the automated detection with a human-in-the-loop inspection of the intermediate representations generated in the previous phase.

To conduct the automated detection, we considered several tools described in the literature (e.g., \textsc{Arcan}~\cite{arcan_tool}, \textsc{MSANose}~\cite{Walker20}, and \textsc{MARS}~\cite{Tighilt2023}). However, we discarded the former ones either because they were not available as open-source or could not be successfully executed in our setting. Ultimately we selected \textsc{MARS}, an open-source tool based 
on static analysis and structural pattern matching, 
capable of identifying 16 anti-patterns that map directly onto the categories defined by Cerny et al.~\cite{Cerny2023MicroserviceAntiPatterns}. 

Although effective, MARS exhibits some limitations with anti-patterns that depend on technology-specific artefacts, such as \textit{No CI/CD} or \textit{No API Gateway}, which may go undetected when projects use alternative frameworks or custom configuration conventions not covered by the tool’s rule set. Likewise, threshold-based detections for \textit{Nano Service} and \textit{Mega Service} are sensitive to domain-specific characteristics: thresholds that are appropriate for enterprise back-end systems may be suboptimal for educational projects or embedded applications, leading to false positives or missed detections.

Therefore, we complemented automated detection with a \textit{human-in-the-loop} analysis. This manual examination of the intermediate representations allowed us to validate tool-reported findings, resolve ambiguous or borderline situations, and detect additional anti-patterns whose identification required contextual understanding of design intent, technology choices, and operational evidence.

The inspection was view-guided: for each anti-pattern category, we analysed the representation(s) most likely to surface meaningful indicators and cross-checked these findings against code, configuration artefacts, and the execution traces collected earlier.

As to the category of \textit{Intra-service Design} anti-patterns, the focus was on each service’s internal structure and responsibilities. Cohesion and responsibility clarity were assessed using the per-service detailed views, simplified internal call graphs (controllers$\rightarrow$services$\rightarrow$repositories), endpoint catalogs, and endpoint, repository mappings, then contextualized with the SonarQube quality table (NCLOC, technical debt) to link ambiguous responsibilities to concrete maintainability concerns.

Turning to the \textit{Inter-service Decomposition} category, the aim was to understand how responsibilities were partitioned and how services depended on one another. The UML Component-and-Connector (C\&C) view provided the high-level topology, interfaces, connectors, and inter-component dependencies, reconstructed from routing/configuration data and observed HTTP traces; endpoint catalogs refined the multi-hop interaction paths visible in the C\&C diagram, revealing symptoms of over-fragmentation and chatty orchestration.

Regarding \textit{Service Interaction}, the emphasis shifted to communication patterns and mediation logic. The C\&C view highlighted unexpected or indirect communication paths, while service and gateway configuration files exposed hardcoded endpoints, missing timeout policies, and a lack of resilience. Execution traces visualized through \textsc{Grafana} dashboards confirmed synchronous call chains and clarified the absence of fallback or retry mechanisms.

From the \textit{Security} anti-pattern perspective, authentication, authorization, exposure, and transport protection were scrutinized by reviewing source code and deployment descriptors (to detect hardcoded secrets and split authentication responsibilities), Docker Compose files and network settings (to identify publicly exposed services and databases), and service configuration (to verify the lack of TLS and access-control enforcement). The combined signals indicated systemic rather than isolated weaknesses.

Lastly, \textit{Team Organization} aspects were inferred from repository-level evidence. The component catalogue reported in Table \ref{tab:components} documented technology choices per component and exposed heterogeneity; the repository contents revealed the presence or absence of CI/CD pipelines; and the logging and monitoring configurations found in code and configuration files revealed the teams’ operational monitoring practices and documentation discipline.

This category-specific, view-guided inspection yielded the curated set of confirmed anti-patterns reported in the Results section, each backed by explicit excerpts from the quality table, the UML C\&C diagram, the per-service detailed models, and the component catalogue.

\subsection{Summary of Results}

The application of the three-phase process led to the identification of 23 out of 58 types of microservice anti-patterns described in \cite{Cerny2023MicroserviceAntiPatterns}. 
In the following, we report, category by category, the observed anti-patterns and the evidence that supports each identification.  For each category, the corresponding tables present the name of the detected anti-pattern along with its numerical identifier according to the taxonomy by Cerny \textit{et al.} \cite{Cerny2023MicroserviceAntiPatterns}, a concise description of its characteristics, and the specific microservices where the anti-pattern was found. When an anti-pattern emerged as a global issue rather than being confined to individual services, we used the label \textit{System-wide}.

\paragraph{\textbf{Intra-service Design}}
According to the results reported in Table \ref{tab:AP_intra_inter}, we identified two occurrences of this category of anti-patterns: 
\begin{itemize}
\item \textit{No API-versioning (n.~6)} system-wide. All externally exposed REST endpoints are published under unversioned URIs (for example, paths of the form \texttt{/api/...} instead of \texttt{/api/v1/...}), and we did not observe any alternative version-negotiation mechanism based on headers or media types. This lack of explicit versioning turns routine refactorings and feature evolution into potentially breaking changes at the system level, and is thus classified as a global instance of the \textit{No API-versioning} anti-pattern. 
\item \textit{Ambiguous service (n.~9)} in \texttt{T23}. The module simultaneously implements player authentication and social-profile management. Co-locating security-critical authentication flows with routine profile CRUD blurs the service boundary, tangles access-control concerns with unrelated functionality, and hinders independent evolution. 
\end{itemize}

Treating identity management and profile features as a single unit of deployment creates unnecessary coupling: changes to social features require retesting authentication paths, and security policies cannot be scoped narrowly, increasing risk and slowing iteration.

\paragraph{\textbf{Inter-service Decomposition}}
Table~\ref{tab:AP_intra_inter} highlights one occurrence in this category:

\begin{itemize}
\item \textit{Chatty Service (n.17)} centered on \texttt{T6}. 
To execute a single game turn initiated by \texttt{T5}, the \textit{Game Engine} (\texttt{T6}) orchestrates several fine-grained, synchronous calls: it queries the \textit{Admin Service} (\texttt{T1}) for the class-under-test code; invokes the \textit{Student Test Runner} (\texttt{T7}) to compile tests and gather coverage; and interacts with the \textit{Game Repository} (\texttt{T4}) to retrieve robot metrics and to update or close the turn. 
The sequence reconstructed from the C\&C view and from the endpoint catalogs reveals an over-fragmented workflow whose success depends on multiple synchronous hops, thereby increasing latency and coupling and elevating the risk of cascading failures. 
Even minor slowdowns in \texttt{T7}, \texttt{T4}, or \texttt{T1} propagate back to \texttt{T6} and ultimately to the user path in \texttt{T5}, illustrating how excessive inter-service chatter amplifies tail latency and reduces resilience.
\end{itemize}

This behavior aligns with the anti-pattern’s definition and reflects a design that favors functional separation over communication efficiency.

\begin{table}[h!]
\label{tab:AP_intra_inter}
\centering
\caption{Intra-service and Inter-service Decomposition Anti-Patterns detected.}
\renewcommand{\arraystretch}{1}
\setlength{\tabcolsep}{6pt}
\setlength{\arrayrulewidth}{0.4pt}

\begin{tabular}{|p{0.18\textwidth}|p{0.50\textwidth}|p{0.10\textwidth}|}
\hline
\centering\bfseries Anti-pattern &
\centering\bfseries Anti-pattern definition &
\centering\bfseries Service(s) \tabularnewline
\hline
\hline
\multicolumn{3}{|c|}{\bfseries Intra-service Design} \tabularnewline
\hline
\hline
\raggedright \textbf{No API-versioning (n.6)} &
\raggedright This anti-pattern occurs when a system exposes its APIs without any versioning strategy, causing connection issues for the API consumers. &
\centering System-wide \tabularnewline
\hline
\raggedright \textbf{Ambiguous service (n.9)} &
\raggedright This anti-pattern occurs when responsibilities of a service are not clearly defined, leading to possible confusion and overlapping functionality. &
\centering T23 \tabularnewline
\hline
\bottomrule
\bottomrule
\multicolumn{3}{|c|}{\bfseries Inter-services Decomposition} \tabularnewline
\hline
\hline
\raggedright \textbf{Chatty service (n.17)} &
\raggedright This anti-pattern arises when a service communicates excessively with other microservices, often performing many fine-grained calls to gather required data. Such over-communication increases response time and degrades overall system performance. &
\centering T6 \tabularnewline
\hline
\end{tabular}
\vspace{4pt}
\captionsetup{justification=centering}
\label{tab:AP_intra_inter}
\end{table}

\paragraph{\textbf{Service Interaction}}
Table~\ref{tab:AP_service_interaction} summarizes the following issues:

\begin{itemize}
\item \textit{On-line only (n.~25)} in \textit{Admin Service} (\texttt{T1}). When an administrator requests test-case generation for a selected class, T1 executes the generation workflow synchronously within the admin UI. The page remains blocked until the job completes, coupling perceived responsiveness to a long-running operation. In the absence of progress indicators or error feedback, user experience degrades and recovery from failures becomes difficult.

\item \textit{Hardcoded endpoint (n.~28)} across \texttt{T1}, \texttt{T5}, and \texttt{T6}. Service locations (IPs and ports) are embedded directly in code or configuration, coupling clients to fixed network coordinates. Such binding complicates scaling and replacement, inhibits service discovery, and increases the risk of deployment drift when environments change.

\item \textit{No API gateway (n.~29)} de facto. Although the \textit{API gateway} exists, most inter-service calls bypass it, preventing centralized mediation for routing, authentication, and policy enforcement. The result is a proliferation of point-to-point integrations that are harder to secure, observe, and evolve coherently. A concrete instance of this anti-pattern can be observed in \texttt{T6}. To execute a turn, \texttt{T6} issues API calls to \texttt{T7} for test compilation and coverage metrics computation, and to \texttt{T4} to update the current game state and record the completed turn. Instead of routing these requests through the \textit{API gateway}, \textit{T6} directly contacts the target microservices.

\item \textit{Wobbly service interactions (n.~30)} system-wide. Calls between services are issued without resilience patterns such as circuit breakers, rate limiting, or fallbacks. Under load or partial failures, this absence makes error propagation more likely and turns transient incidents into visible outages.

\item \textit{Timeout (n.~31)} system-wide. Client requests (e.g., \texttt{T1}$\rightarrow$\texttt{T4}, \texttt{T6}$\rightarrow$\texttt{T7}/\texttt{T4}) do not specify explicit time limits, allowing calls to hang indefinitely or fail unpredictably. Without bounded waiting, upstream services become vulnerable to resource starvation and cascading slowdowns.
\end{itemize}

\begin{table}[h!]
\centering
\caption{Service Interaction Anti-Patterns detected.}

\renewcommand{\arraystretch}{1}
\setlength{\tabcolsep}{6pt}
\setlength{\arrayrulewidth}{0.4pt}

\begin{tabular}{|p{0.18\textwidth}|p{0.59\textwidth}|p{0.10\textwidth}|}
\hline
\centering\bfseries Anti-pattern &
\centering\bfseries Anti-pattern definition &
\centering\bfseries Service(s) \tabularnewline
\hline
\multicolumn{3}{|c|}{\bfseries Service Interaction} \tabularnewline
\hline
\hline
\raggedright \textbf{On-line only (n.25)} &
\raggedright This anti-pattern occurs when a system relies entirely on synchronous communication between microservices and avoids using batch processing, even for tasks that are long-running or resource-intensive, leading to performance bottlenecks and inefficiencies. &
\centering T1 \tabularnewline
\hline
\raggedright \textbf{Hardcoded endpoint (n.28)} &
\raggedright This anti-pattern occurs when microservice addresses, ports, or endpoints are directly embedded in code, configuration, or environment variables, preventing, for example, easy scaling. &
\centering T1, T5, T6 \tabularnewline
\hline
\raggedright \textbf{No API-gateway (n.29)} &
\raggedright This anti-pattern occurs when a microservice system lacks a centralized API gateway, forcing clients to call services directly, complicating various aspects such as security and routing. &
\centering System-wide \tabularnewline
\hline
\raggedright \textbf{Wobbly service interactions (n.30)} &
\raggedright This anti-pattern occurs when a service interacts with other services or message routers without mechanisms to tolerate failures. &
\centering System-wide \tabularnewline
\hline
\raggedright \textbf{Timeout (n.31)} &
\raggedright This anti-pattern occurs when a service consumer frequently fails to connect to a microservice, possibly due to unhandled service unavailability. &
\centering System-wide \tabularnewline
\hline
\end{tabular}

\vspace{4pt}
\captionsetup{justification=centering}

\label{tab:AP_service_interaction}
\end{table}

\paragraph{\textbf{Security}}
Table~\ref{tab:AP_security} highlights a pervasive lack of foundational safeguards. The following issues were observed:

\begin{itemize}
\item \textit{Unauthenticated traffic (n.~33)} system-wide. Internal APIs accept requests without verifying the caller’s identity. In practice, any reachable service, potentially even an external client if a port is exposed, can invoke privileged operations, eroding trust boundaries and enabling lateral movement.

\item \textit{Multiple user authentication (n.~34).} Authentication is split between \texttt{T1} (administrators) and \texttt{T23} (players). This fragmentation enlarges the attack surface, makes policy drift more likely, and complicates consistent enforcement of password, session, and token rules across the platform.

\item \textit{Publicly accessible microservices (n.~35).}  In the \texttt{Docker Compose} configurations of services \texttt{T1}, \texttt{T4}, \texttt{T23}, and \texttt{T8}, the databases embedded within these microservices were exposed publicly through the ports directive, which mapped internal database ports directly to the host. This configuration made the databases accessible outside the Docker network, bypassing the intended gateway and centralized access controls. As a result, internal data stores were directly reachable from external networks, revealing implementation details and significantly increasing the system’s attack surface.

\item \textit{Unnecessary privileges (n.~36)} and \textit{Insufficient access control (n.~37)} system-wide. 
Each microservice can invoke all the functionalities of every other microservice since the system does not implement authentication or authorization. In effect, the absence of authentication implies the absence of access control, granting unlimited, cross-service access. The problem is further enhanced by predominantly direct service-to-service calls that bypass the API gateway, eliminating a central point of control for authorization checks or policy enforcement.

\item \textit{Non-secured service-to-service communications (n.~39)} system-wide. Links between microservices are unencrypted (no TLS), leaving data in transit vulnerable to interception and tampering, and preventing mutual authentication of peers.

\item \textit{Hardcoded secrets (n.~42).} Credentials and other sensitive tokens appear directly in source or deployment scripts for \texttt{T1}, \texttt{T23}, and \texttt{T4}. Such embedding promotes leakage through version control, logs, or error messages and obstructs rotation and revocation practices.
\end{itemize}

Public exposure of service and database ports, combined with unauthenticated, non-TLS traffic and hardcoded credentials, forms a direct path to data exfiltration and unauthorized manipulation. Rather than isolated mistakes, these co-occurring smells indicate structural security debt with immediate consequences for confidentiality, integrity, and operational resilience.

\begin{table}[h!]
\centering
\caption{Security Anti-Patterns detected.}
\renewcommand{\arraystretch}{1}
\setlength{\tabcolsep}{6pt}
\setlength{\arrayrulewidth}{0.4pt}

\begin{tabular}{|p{0.22\textwidth}|p{0.55\textwidth}|p{0.10\textwidth}|}
\hline
\centering\bfseries Anti-pattern &
\centering\bfseries Anti-pattern definition &
\centering\bfseries Service(s) \tabularnewline
\hline
\hline
\multicolumn{3}{|c|}{\bfseries Security} \tabularnewline
\hline
\raggedright \textbf{Unauthenticated traffic (n.33)} &
\raggedright This anti-pattern occurs when API requests, either from external clients or between internal microservices, are not properly authenticated. &
\centering System-wide \tabularnewline
\hline
\raggedright \textbf{Multiple user authentication (n.34)} &
\raggedright This anti-pattern occurs when a microservice system provides multiple access points for user authentication, increasing the attack surface and potential security risks. &
\centering T1, T23 \tabularnewline
\hline
\raggedright \textbf{Publicly accessible microservices (n.35)} &
\raggedright This anti-pattern occurs when microservices are directly exposed to external clients. Exposing services without proper control can reveal sensitive information and increase the risk of attacks. &
\centering T1, T23, T4, T8 \tabularnewline
\hline
\raggedright \textbf{Unnecessary privileges to microservices (n.36)} &
\raggedright This anti-pattern occurs when microservices are granted more access, permissions, or functionalities than necessary to perform their tasks. &
\centering System-wide \tabularnewline
\hline
\raggedright \textbf{Insufficient access control (n.37)} &
\raggedright This anti-pattern occurs when microservices lack proper access control, potentially exposing sensitive data or business functions and violating confidentiality. &
\centering System-wide \tabularnewline
\hline
\raggedright \textbf{Non-secured service-to-service communications (n.39)} &
\raggedright This anti-pattern occurs when service-to-service interactions are not encrypted or secured. This exposes data to interception, tampering, and eavesdropping. &
\centering System-wide \tabularnewline
\hline
\raggedright \textbf{Hardcoded secrets (n.42)} &
\raggedright This anti-pattern occurs when sensitive configuration data, such as passwords or API keys, are embedded directly in source code or deployment scripts. &
\centering T1, T23, T4 \tabularnewline
\hline
\end{tabular}

\vspace{4pt}
\captionsetup{justification=centering}

\label{tab:AP_security}
\end{table}

\paragraph{\textbf{Team organization}}

Table~\ref{tab:AP_team_org} surfaces organizational anti-patterns that reflect process gaps and governance weaknesses:

\begin{itemize}
\item \textit{Golden hammer (n.~44)} system-wide. Students consistently selected technologies based on prior familiarity or personal preference rather than fitness for purpose. For example, the \textit{Game Repository} (\texttt{T4}) was implemented in \texttt{Go} even though most of the system had already been built in \texttt{Java}, because the student team that developed \texttt{T4} had prior experience with \texttt{Go}. 

\item \textit{Too many standards (n.~46)} system-wide. The system shows marked heterogeneity: \texttt{T4} is implemented in \texttt{Go}, \texttt{T8} in \texttt{Node.js}, while most other services are Java-based; databases span \texttt{MongoDB} (\texttt{T1}), \texttt{MySQL} (\texttt{T23}), and \texttt{PostgreSQL} (\texttt{T4}). 
Diversity can be pedagogically useful, but ungoverned fragmentation inflates onboarding time, complicates integration across language/runtime boundaries, and raises operational overhead.

\item \textit{Inadequate techniques support (n.~47)} system-wide.  Uneven skill levels led to inconsistent toolchains and ad-hoc library choices. 
The student teams solved similar problems in divergent ways, reducing reuse and making system-wide improvements difficult to propagate. This fragmentation was evident in the recovered architectural model: cross-cutting concerns such as authentication, logging, and communication protocols were inconsistently handled across services. Several teams reimplemented gateway logic or direct REST integrations independently, leading to duplicated effort, inconsistent interfaces, and limited reuse of common infrastructure. 

\item \textit{No CI/CD (n.~53)} system-wide. The absence of continuous integration and delivery removed a key feedback loop. Changes were integrated manually, delaying defect discovery, discouraging small, frequent releases, and increasing the likelihood of configuration drift across environments.

\item \textit{Manual Configuration (n.~54)} system-wide. Service configuration was maintained through manual edits to per-service \texttt{.properties}/\texttt{.yml} files, \texttt{.env} snippets, and Docker Compose descriptors, with credentials, hostnames, and ports duplicated across multiple artifacts. No dedicated configuration service or external configuration management tool was used. As a result, automated roll-outs and environment-specific provisioning were not supported: deploying or scaling the system required developers to modify several files by hand, increasing the risk of inconsistent settings, hard-coded secrets, and configuration drift.

\item \textit{Insufficient monitoring (n.~55)} with \textit{Local logging (n.~58).} Each service logged in isolation with minimal structure and no central aggregation. Lacking metrics and traces, teams had little visibility into performance regressions or failure propagation, making diagnosis slow and incident response reactive rather than proactive.

\item \textit{Dismiss documentation (n.~56)} system-wide. The project lacked a unified documentation style and governance. Most REST endpoints had no up-to-date OpenAPI/Swagger specifications; payload schemas, error models, and auth flows were undocumented or inconsistently described. Without a single source of truth or style rules (naming, versioning, examples), teams relied on implicit contracts and ad-hoc notes, leading to divergent conventions, brittle integrations, and higher change costs as assumptions accumulated outside versioned specs.

\end{itemize}

\begin{table}[h!]
\centering
\caption{Team Organization Anti-Patterns detected.}
wcommand{\arraystretch}{1}
\setlength{\tabcolsep}{6pt}
\setlength{\arrayrulewidth}{0.4pt}

\begin{tabular}{|p{0.18\textwidth}|p{0.58\textwidth}|p{0.10\textwidth}|}
\hline
\centering\bfseries Anti-pattern &
\centering\bfseries Anti-pattern definition &
\centering\bfseries Service(s) \tabularnewline
\hline
\multicolumn{3}{|c|}{\bfseries Team organization} \tabularnewline
\hline
\hline
\raggedright \textbf{Golden hammer\\ (n.44)} &
\raggedright This anti-pattern occurs when a familiar technology is applied to every problem, regardless of whether it is appropriate. It often arises when developers repeatedly reuse a technology they know well, even for problems that do not require it. &
\centering System-wide \tabularnewline
\hline
\raggedright \textbf{Too many\\ standards (n.46)} &
\raggedright This anti-pattern occurs when multiple development languages, protocols, or frameworks are used across microservices. While microservices allow technology diversity, excessive heterogeneity can create maintenance challenges, especially with developer turnover, and complicate integration and long-term evolution. &
\centering T1, T23, T4, T8 \tabularnewline
\hline
\raggedright \textbf{Inadequate\\ techniques\\ support (n.47)} &
\raggedright This anti-pattern occurs when the development team uses tools, processes, or methods that are ill-suited for building or managing microservices. Such choices, often made despite available alternatives, can introduce disadvantages and hinder effective development and operation. &
\centering System-wide \tabularnewline
\hline
\raggedright \textbf{No CI/CD\\ (n.53)} &
\raggedright This anti-pattern occurs when continuous integration and deployment practices are not used. &
\centering System-wide \tabularnewline
\hline
\raggedright \textbf{Manual Configuration (n.54)} &
\raggedright This anti-pattern occurs when hosts, services, and instances are configured manually, preventing a clear separation between the core codebase and configuration management to enable automation. &
\centering System-wide \tabularnewline
\hline
\raggedright \textbf{Insufficient\\ monitoring (n.55)\\ and Local logging\\ (n.58)} &
\raggedright Lack of monitoring occurs when microservices’ performance and failures are not tracked, making issues harder to detect and resolve; Local logging occurs when each microservice manages logs independently, without centralized aggregation or monitoring. &
\centering System-wide \tabularnewline
\hline
\raggedright \textbf{Dismiss\\ documentation\\ (n.56)} &
\raggedright This anti-pattern occurs when exposed APIs lack proper documentation, making it difficult to maintain an overall system view. &
\centering System-wide \tabularnewline
\hline
\end{tabular}

\vspace{4pt}
\captionsetup{justification=centering}

\label{tab:AP_team_org}
\end{table}

\begin{figure}[h!]
  \centering
  \includegraphics[width=0.8\linewidth]{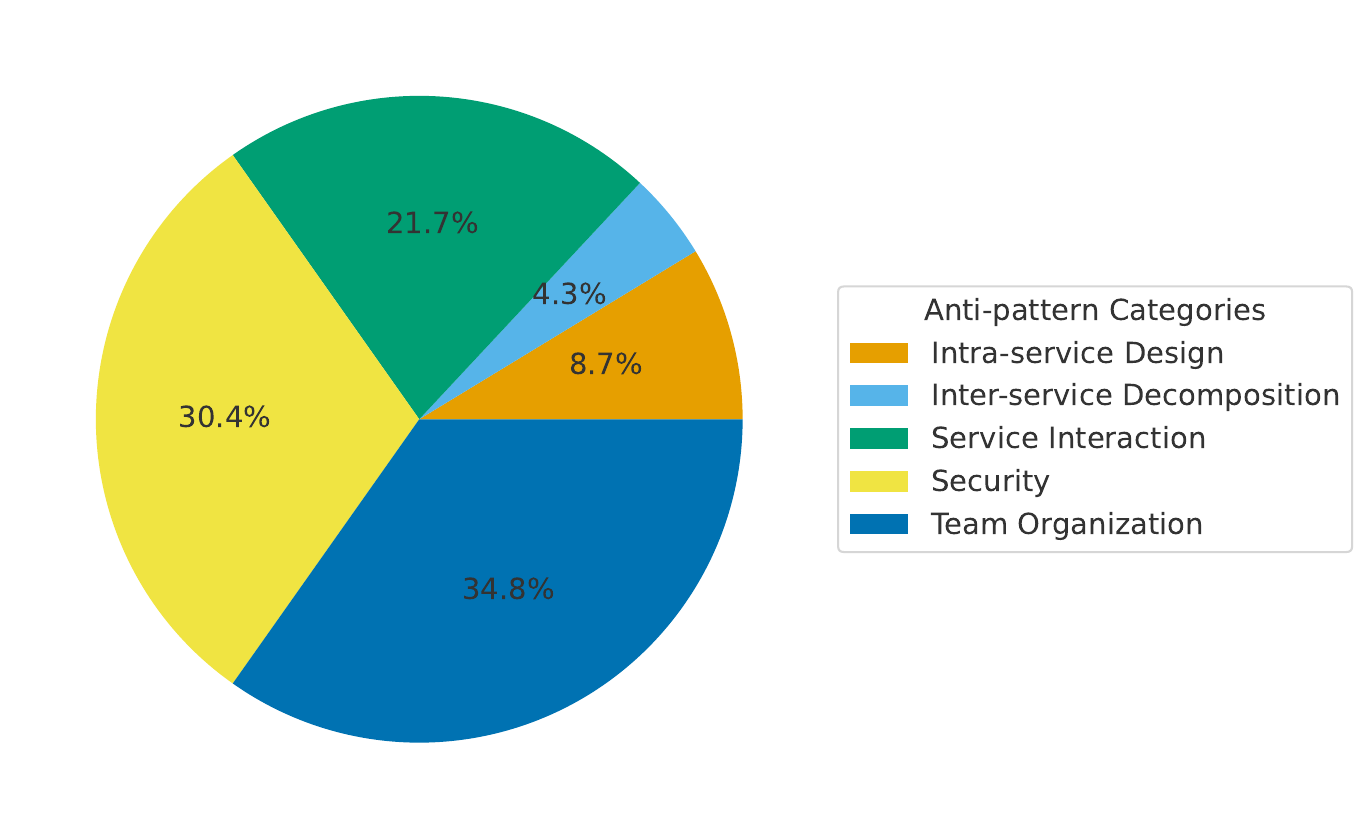}
  \caption{Distribution of detected Anti-patterns across the five analyzed categories.}
  \label{fig:ap_pie}
\end{figure}

Figure \ref{fig:ap_pie} shows the distribution of detected anti-patterns across the five analyzed categories.  The Security category was the most frequently observed (34.8\%), indicating recurring challenges in implementing authentication, authorization, and data protection mechanisms. The Team Organization category (30.4\%) followed, mainly reflecting students’ limited experience with DevOps practices and collaborative workflows. The Service Interaction category (21.7\%) came next, revealing common issues in inter-service communication and coordination. Fewer anti-patterns were found in Inter-service Decomposition (4.3\%) and Intra-service Design (8.7\%), suggesting that students generally managed service boundaries and internal structures effectively.

\color{black}

\section{Discussion}
\label{sec:discussion}

This section discusses the results of the three-phase experimental procedure in relation to our five research questions (RQs). Each subsection interprets the findings presented in Section~\ref{sec:Results}, connecting them to existing literature and reflecting on their implications for microservice architecture education.

\subsection{Answering RQ1: Which ``\textit{Intra-service Design}'' anti-patterns occur in  MSAs  developed by students?}

This category refers to flaws in the internal design of individual microservices that can result in poorly sized or weakly cohesive components, such as services that combine too many or too few responsibilities, or expose functionality that is unclear and not well-aligned with a specific purpose.

As reported in Table~\ref{tab:AP_intra_inter}, we identified only \textbf{2 out of 9} possible intra-service anti-patterns. Specifically, we found a single instance of \textit{Ambiguous Service} in \texttt{T23} and a system-wide occurrence of \textit{No API-versioning}.
In contrast, we did not observe other common intra-service issues such as overly fine-grained services with few operations (\textit{Nano-service}), large and multifaceted services (\textit{Mega-services}), services containing unused functionality (\textit{Nobody home}), or interfaces exposing full CRUD operations at the service boundary (\textit{CRUDY service}).

This limited occurrence suggests that students generally have a solid understanding of internal microservice design. The lone exception involved bundling a cross-cutting concern (i.e., user account and authentication) with feature-specific functionality. Analysis of \texttt{T23}'s evolution suggests that, under time pressure and with evaluation criteria favoring demonstrable features, students may have chosen to extend an existing service as a quicker and safer approach rather than designing a separate, well-bounded component.

Notably, the \textit{Ambiguous Service} anti-pattern found in the students' projects has also been reported as the third most frequent service anti-pattern in the systematic literature review on smell detection presented by \cite{Sabir2019}. This fact indicates that the tendency to design weakly cohesive services is not limited to novice developers but is instead a recurring issue observed across diverse software engineering contexts.
It suggests that developers, regardless of experience level, often struggle to balance service granularity, highlighting the need for clearer guidelines, better tool support, and more effective educational strategies to help practitioners avoid creating overly fine-grained services.

As to the second anti-pattern detected, the system-wide occurrence of \textit{No API-versioning} showed insufficient attention of the students to the necessity of long-term API maintenance and evolution. This lack could be due to how the topic was covered in the course: API versioning was not emphasized enough as a mandatory design practice, so most groups retained the initial, unversioned endpoint structure introduced during early prototyping and did not revise it.

\begin{tcolorbox}[colback=black!5, colframe=black!75, coltitle=white, title=Educational Finding]
The occurrence of the \textit{Ambiguous Service} and \textit{No API-versioning} anti-patterns indicates that, even when students are familiar with the basic principles of microservice architectures, they may struggle to consistently apply service decomposition and intra-service design, as well as software evolution and maintenance practices.
Teaching activities should therefore emphasize hands-on exercises on service boundaries, modularity, and the long-term impact of design decisions, helping students understand how internal service design affects the overall quality and maintainability of the system.
\end{tcolorbox}

\subsection{Answering RQ2: Which ``\textit{Inter-service Decomposition}'' anti-patterns occur in MSA architectures developed by students?}

This category concerns flaws in the division of functionality across multiple microservices, such as duplicated responsibilities, cyclic call dependencies, or excessive inter-service communication to complete a task.

As reported in Table~\ref{tab:AP_intra_inter}, 
there was a limited occurrence of inter-service anti-patterns in the analyzed architecture. We observed only \textbf{1 out of 14} anti-pattern of this category, namely \textit{Chatty Service} in \texttt{T6}, the Game Engine service. 
No instances were found instead of unnecessary service proliferation (\textit{Microservice greedy}), cyclic dependencies (\textit{Cyclic dependencies}), or layer-based splits that violate cohesive responsibilities (\textit{Wrong cuts}).

Regarding the \textit{Chatty Service}, its role as the Game Engine inherently required multiple interactions with other microservices, making part of this behavior expected for an orchestrator.
However, the absence of an asynchronous communication mechanism (e.g., a message broker) in the architecture, likely due to time constraints or limited experience, forced students to rely solely on synchronous REST calls, amplifying the appearance of this anti-pattern.

With respect to the frequency of service anti-patterns reported in the literature review by \cite{Sabir2019}, it must be noted that the \textit{Chatty Service} was also among the most frequently observed, appearing as the fourth most commonly reported one. 
This finding suggests that excessive inter-service communication is a persistent challenge in microservice design, affecting not only novice developers but also practitioners more broadly.

\begin{tcolorbox}[colback=black!5, colframe=black!75, coltitle=white, title=Educational Finding]
This case highlights a slight tendency toward over-decomposition and reliance on fine-grained synchronous interactions, leading to a \textit{Chatty Service}. To mitigate similar issues, teaching activities should emphasize design reasoning on service granularity, hands-on practice with asynchronous communication (e.g., event-driven design, sagas, message queues), and early design reviews that visualize inter-service dependencies and discuss latency and communication trade-offs. Such activities can help students internalize the balance between cohesion and communication cost in microservice architectures.
\end{tcolorbox}

\subsection{Answering RQ3: Which ``\textit{Service Interaction}'' anti-patterns occur in MSAs  developed by students?}

Service interaction anti-patterns arise from inefficient or improper communication patterns between microservices, often leading to increased latency, tighter coupling, or failure propagation across the system.

As reported in Table~\ref{tab:AP_service_interaction}, we identified \textbf{5 out of 9} anti-pattern types belonging to this category. Among these, two (\textit{Hardcoded Endpoints} and \textit{On-line Only}) were found in three services, while two others (\textit{Wobbly Service Interactions} and \textit{Timeout}) affected the entire architecture, as they appeared in nearly all microservices. The fifth anti-pattern, \textit{No API Gateway}, was detected at the system level, since no component consistently leveraged the available gateway for outgoing calls.

These findings suggest that students were able to implement functional communication between services, but often overlooked robustness and adaptability aspects of inter-service interactions. The presence of \textit{Hardcoded Endpoints} and the absence of systematic gateway usage indicate limited attention to configuration management and request routing flexibility, which are key to maintainability and scalability in microservice systems. Similarly, the occurrence of \textit{Wobbly Service Interactions} and \textit{Timeout} issues suggests a lack of resilience mechanisms such as retry policies, circuit breakers, or asynchronous communication strategies.

A plausible explanation is that, under strong time constraints and an evaluation focus on delivering functional features, students prioritized “\textit{getting things to work}” over designing resilient and maintainable interaction patterns. 
This behavior reflects a common tendency among novice developers to treat inter-service communication as a secondary concern rather than as a first-class architectural element. Similar issues also emerge in professional settings, where practitioners often do not rely on architectural tools or metrics, even when dealing with tasks such as service decomposition and microservice integration, which are inherently architectural in nature~
 
Moreover, several anti-patterns identified in the students’ projects, such as \textit{Hardcoded Endpoint} and \textit{No API Gateway}, have likewise been reported by practitioners as among the most severe and risky~\cite{8354414}. This alignment between academic and industrial observations underscores the persistent difficulty of managing architectural concerns in microservice systems and highlights the need for improved design guidance and more focused educational experiences.

\begin{tcolorbox}[colback=black!5, colframe=black!75, coltitle=white, title=Educational Finding]
This case reveals a tendency to prioritize functionality over resilience, leading to fragile communication patterns such as \textit{Hardcoded Endpoints} and \textit{Timeouts}. To address this, future teaching activities should emphasize resilient communication design, including hands-on use of API gateways, service discovery, and fault-tolerance mechanisms (e.g., retries, circuit breakers, backoff strategies). Introducing asynchronous or event-driven interaction exercises can also help students internalize scalability and fault isolation principles essential for robust microservice systems.
\end{tcolorbox}

\subsection{Answering RQ4: Which ``\textit{Security}'' anti-patterns did  occur in  MSAs developed by students?}

This category concerns flaws in authentication, authorization, and secure service-to-service communication. Typical violations include unauthenticated service-to-service calls, insufficient access control or lack of central enforcement, hardcoded secrets, and exposure of internal services to the public network.

As summarized in Table~\ref{tab:AP_security}, we observed \textbf{7 out of 10} security anti-pattern types. Most occurred at the system-wide level: \textit{Unauthenticated traffic}, \textit{Insufficient access control}, \textit{Non-secured service-to-service communications}, and \textit{Unnecessary privileges to microservices} were present across the architecture. In addition, \textit{Multiple user authentication} appeared in two services, \textit{Hardcoded secrets} in three services, and \textit{Publicly accessible microservices} in five services.

A plausible explanation is that teams prioritized functional delivery under time constraints and relied on default configurations, implicitly treating the internal network as trusted. Limited familiarity with core security practices, centralized authentication and authorization, least-privilege enforcement, secret management, and service-to-service encryption made ad-hoc solutions appear acceptable during local development and demos. Docker-based setups with fixed port mappings further normalized exposing internal services, while the absence of mandatory security checks (timeouts for authentication flows, token validation, TLS by default) reduced early feedback on these choices.

However, these results were largely expected: students did not have a strong prior background in security, and the topic was outside the scope of the course. 
At the same time, it is noteworthy that the issues they introduced align with well-known security smells extensively documented by researchers and practitioners \cite{ponce2022smells}. This convergence suggests that such problems are not merely the result of limited instruction but reflect broader, recurring challenges in secure microservice design. The findings, therefore, highlight curricular boundaries while also pointing to meaningful opportunities to strengthen instructional coverage of security-related architectural concerns.

\begin{tcolorbox}[colback=black!5, colframe=black!75, coltitle=white, title=Educational Finding]
Security weaknesses were widespread, consistent with students’ limited background and the course scope. Future offerings should integrate security-by-design basics for microservices: centralized authentication and authorization, least privilege and scoped roles, secret management instead of hardcoded credentials, and encrypted service-to-service communication (preferably mutual TLS). Include hands-on labs that secure a baseline stack (gateway or service mesh, discovery, default-deny networking) and make these controls part of the assessment, reflecting the centrality of security in modern, widely adopted microservice architectures.
\end{tcolorbox}

\subsection{Answering RQ5: Which ``\textit{Team organization}'' anti-patterns occur in  MSAs  developed by students?}

This category concerns both flaws arising from wrong practices and standards adopted by development teams, and operational problems deriving from violations of DevOps
practices. 
Typical problems of the former type include missing or excessive standards for service communication, whereas the latter comprises the lack of CI/CD, and manual configuration. 

As summarized in Table~\ref{tab:AP_team_org}, we observed \textbf{8 out of 16} anti-pattern types in this category.
Consistent with the definition of this category, focused on development practices rather than structural properties of individual services, we report team-organization anti-patterns as \textit{system-wide} findings. Attributing them to single services would be misleading, since they stem from shared process and platform choices applied across the whole architecture.

In our case, they manifested as technology choices driven by familiarity (\textit{Golden hammer}), excessive heterogeneity across stacks (\textit{Too many standards}), uneven toolchains and ad-hoc library use (\textit{Inadequate technique support}), absence of continuous integration and delivery (\textit{No CI/CD}), manual management of configuration artifacts across services (\textit{Manual configuration}), limited observability (\textit{Insufficient monitoring} and \textit{Local logging}), and lack of up-to-date, authoritative API documentation (\textit{Dismiss documentation}).

A plausible didactic explanation is that teams tended to select technologies with which they already had prior familiarity (from coursework or personal projects). In the absence of guardrails, this familiarity bias reinforced heterogeneity: students’ technology choices reflected prior familiarity more than problem fit, leading teams to prefer known stacks over a common baseline and to proceed without a structured comparison of candidate solutions. 
Limited students' preparation and practice in DevOps, combined with missing starter templates and pipelines, made automation optional. Moreover, the limited emphasis on logs/metrics/traces and documentation further reduced incentives to invest in observability and common standards.
These team organization anti-patterns are common even in projects developed by professionals, as reported by \cite{Taibi2019MicroservicesAntiPatterns, AgostiniSB24, ci_bad_practices, ZhouHZHSZ22}.

\begin{tcolorbox}[colback=black!5, colframe=black!75, coltitle=white, title=Educational Finding]
These findings point to process and platform gaps rather than conceptual misunderstandings of microservices. Instructors should provide a golden path with minimal, enforced standards (service communication conventions, API contracts), a ready-to-use CI/CD starter kit, a shared configuration management solution, and baseline observability (logs, metrics, traces) so that automation and monitoring are the default. Strengthening practices such as enforcing authoritative API documentation, limiting uncontrolled stack heterogeneity, and avoiding manual, service-specific configuration can help students better understand how organizational decisions influence the evolvability and operational reliability of microservice systems.
\end{tcolorbox}

\subsection{Threats to Validity}

We acknowledge several potential threats to the validity of our study and describe how we attempted to mitigate them.
    
\textbf{Internal validity.} 
Internal validity concerns whether the observed results are actually caused by the investigated factors rather than by external variables. In our study, participants were novice programmers by design, as the goal was to analyze anti-patterns emerging in educational settings. A potential threat could stem from differences in students’ prior experience, motivation, or teamwork dynamics. To assess this risk, we collected data on students’ backgrounds and found a high degree of homogeneity in programming experience and familiarity with the technologies involved, which mitigates the impact of individual variability. Moreover, all teams worked under the same requirements, technological stack, and tutoring conditions, further reducing uncontrolled differences. Some variability in performance may still remain, and future replications with cohorts from different contexts or academic years would help strengthen internal validity.

Another internal validity threat is that some observed anti-patterns may stem from the instructional design itself rather than from students’ intrinsic difficulties. To mitigate this risk, the course structure, examples, and guidelines were aligned with well-established textbooks and widely adopted best practices in software engineering. Moreover, all students received the same materials and tutoring support, reducing the likelihood that systematic mistakes originated from the teaching approach.  To further mitigate this threat, future studies should compare outcomes across different instructors or teaching styles, or triangulate these findings with data from parallel courses or different institutions.

A further internal validity threat concerns the fact that our analysis was conducted on a single, integrated version of the project, built from a subset of student implementations selected for completeness and ease of integration. While this ensured a consistent baseline across course editions, it may have introduced selection bias by excluding more heterogeneous or problematic solutions, potentially affecting the distribution of observed anti-patterns.
To mitigate this threat, in future work, we plan to replicate our teaching experience with different projects in order to better capture the diversity of student implementations.
Replicating the analysis across different cohorts or institutions will further strengthen internal validity.

\textbf{Construct validity.} 
Construct validity concerns whether the study effectively captures the intended concepts. In our case, this relates to the accuracy of mapping the detected issues to recognized microservice anti-patterns. Although the classification was grounded in established taxonomies \cite{Cerny2023MicroserviceAntiPatterns}, some design choices made by students may still admit multiple interpretations or only partially align with canonical definitions.

To reduce subjectivity, multiple researchers independently reviewed the identified issues and assigned anti-pattern labels using a shared evaluation protocol. Disagreements were resolved through discussion until consensus was achieved, increasing the consistency of the classification. Each label was also cross-checked against the formal definitions in the reference taxonomy to ensure conceptual fidelity.

\textbf{External validity.}
External validity concerns the extent to which our findings can be generalized beyond the specific context of this study. Our analysis focuses on a single educational project; therefore, the results cannot be directly generalized to other instructional settings or to professional development contexts.
However, the project can be considered reasonably representative within educational environments, as it incorporates contributions from more than 200 students across three academic years, providing a diverse range of design and implementation decisions.
While this breadth does not eliminate limitations to generalizability, it does strengthen the relevance of the observed patterns within similar educational scenarios.

To further mitigate this threat, additional studies conducted in different institutions, with varied pedagogical settings and heterogeneous application domains, are needed to validate and extend our findings.

\textbf{Conclusion validity.} 
Conclusion validity refers to the reliability of the relationships and conclusions drawn from the data. 
Given the qualitative nature of the analysis, conclusions are interpretative rather than statistical. 
We mitigated this threat by systematically mapping each identified issue to a known anti-pattern category and grounding our reasoning in established literature.

\section{Related Works}
\label{sec:Related}

For the related work discussed in this paper, we consider three main research areas. The first encompasses studies that investigate real-world microservice architectures with respect to their quality and the presence of architectural anti-patterns. 
The second covers work reporting experiences and methodologies for teaching microservice-based architectures. The third area focuses on research that examines the quality of student-developed software projects, particularly in educational settings.

\subsection{Studies on MSA quality and anti-patterns }

Several empirical studies have investigated microservice architecture (MSA) anti-patterns, either by inspecting real-world systems or by surveying practitioners.

Tighilt et al. \cite{Tighilt2020} conducted a systematic review and analyzed 67 open-source microservice systems, deriving a catalog of 16 microservice anti-patterns from their empirical observations.
Years later, the same authors \cite{Tighilt2023} proposed MARS (Microservice Anti-patterns Research Software), a fully automated approach supported by a framework for specifying and identifying microservice anti-patterns. Using MARS, the authors specified and detected the 16 anti-patterns across 24 microservice-based systems.

Bakhtin et al. \cite{Bakhtin2025} investigated the use of graph-specific metrics, such as network centrality, to assess MSA quality and detect architectural anti-patterns. Their study involved 24 open-source MSA projects comprising 53 microservices.

Dynamic and runtime analyses have also been explored to detect anti-patterns that static inspection may miss. For example, Parker et al. \cite{Parker2023} conducted a systematic mapping study to characterize the current state of research on detecting microservice anti-patterns from a dynamic perspective.
Finally, several empirical studies have examined how structural anti-patterns correlate with performance and scalability issues in open-source case studies and benchmark systems \cite{Avritzer2025}.

A second research branch elicits practitioners’ knowledge to characterize how widespread and impactful microservice anti-patterns are in industry. Taibi et al.  \cite{Taibi2019MicroservicesAntiPatterns} built a taxonomy of microservice anti-patterns grounded in practitioner interviews and experience, producing a practitioner-informed catalog of technical and organizational anti-patterns.
Cerny et al. \cite{Cerny2023MicroserviceAntiPatterns} conducted a tertiary study synthesizing evidence from surveys and interviews, revealing gaps between academic detection techniques and the issues practitioners report in real projects.
Taibi and Lenarduzzi \cite{8354414} interviewed 72 experienced developers over two years, focusing on bad practices encountered during the development of microservice-based systems. Their study resulted in a catalog of 11 microservice-specific bad smells ranked by perceived harmfulness.

These practitioner-centered works are important because they complement static and dynamic detection approaches with evidence about real-world prevalence, root causes (e.g., organizational or process-related factors), and feasible refactorings.
Nevertheless, they fail to consider the educational context and do not explore the architectural quality of microservice-based systems implemented by students.

\subsection{Microservice architecture education}

There are still very few scientific studies on teaching microservice-based software architectures.

A first group of studies focuses on software architecture education in general, covering methods, materials, role-playing, and simulations, but not specifically on microservices. For example, Meissner et al. \cite{Meissner2025} report on their project-based learning approach that incorporates gamification to improve motivation in collaborative software project development within an agile process context. In this approach, collaboration is limited to the members of a single student group, each playing different roles (e.g., customer, Scrum Master, developers). In contrast, our approach considers collaborative development across multiple student groups.

Another category includes works investigating applications of microservices in educational platforms, focusing more on using the architecture rather than teaching it. For example, Meissner and Thor (2021) \cite{Meissner2021EAsLiT} propose EAs.LiT v2, an adaptable educational platform built on microservices, highlighting the flexibility and scalability of microservice-based educational tools.

A third branch comprises publications describing practical experiences in teaching microservices through case studies, DevOps projects, or specific technological stacks. Examples include Ferreira et al. (2025) \cite{Ferreira2025TeachingMS}, who describe teaching microservices via industry-like teamwork with undergraduate students. Another study reports experiences where students refactored a monolithic application into microservices while adopting DevOps practices \cite{TeachingMSADevOps2022}. In the context of curriculum reform based on Outcome-Based Education (OBE), Qian (2025) \cite{Qian2025OBEMicroservices} presents a project-based teaching model for Spring Cloud microservices, which improved student learning outcomes.

A study closely related to ours is by Lau et al. \cite{Lau2024}, who propose a Collaborative Software Development Project Framework for a course aimed at teaching microservice architectures. This study differs from ours in that it does not focus on the quality of students’ projects. Instead, it investigates students’ ability to connect theory with practice by assessing whether they can identify and address real-world problems, apply technical skills acquired during the course, develop social and cognitive competencies through project work, and critically reflect on outcomes to identify opportunities for improvement.

Although these contributions provide valuable insights into teaching microservice architectures, none specifically investigate the quality of microservice projects developed by students. Moreover, they do not address realistic student projects involving multiple teams working across several development iterations.

\subsection{Studies on Software Quality of Student-Developed Projects}

Several works in the literature focus on analyzing the quality of student projects for different educational aims. 

Many contributions primarily focus on supporting teachers through the automatic assessment of student-developed software quality. 
The systematic literature review reported in \cite{Keuning2018} shows that more than one hundred automated assessment tools have been developed, providing a variety of functions across multiple programming languages and execution environments. More recently, Chen et al. \cite{Chen2022} proposed an Automated Programming Assessment System (APAS) featuring a code-quality evaluation scheme designed to overcome difficulties in assessing the contributions of individual team members.

Another relevant line of research investigates the quality of student-developed projects with the goal of enhancing teaching practices. Keuning et al. \cite{Keuning2017} examined the presence of quality issues in a large set of student Java programs, exploring which issues occur most frequently, whether students are able to resolve them over time, and whether the use of code analysis tools influences issue occurrence. Hamer et al. \cite{Hamer2021} conducted a case study analyzing Git repositories to explore whether students’ changes affect project quality, identifying challenges and improvement opportunities in their engineering practices. 
More recently, De Luca et al. \cite{Deluca2024} implemented a pipeline based on advanced frameworks such as ArchUnit and SonarQube, demonstrating its usefulness as a resource for educators to provide concrete evidence of quality problems in student projects. The same authors in 2025 \cite{DeLuca2025} observed that software quality in student projects is often overlooked, yet systematic analyses of Java-based assignments can provide actionable feedback to enhance software engineering courses. Similarly, Di Meglio et al. (2025) \cite{DiMeglio2025} investigated design rule violations in REST APIs of web applications developed by students. Their findings highlight the need for more focused instruction on API design and support the adoption of automated tools to improve code quality in student projects.

Despite the breadth of these studies, none have specifically addressed the architectural quality of microservice-based systems developed by students, leaving this aspect largely unexplored in the literature.

\section{Conclusions and Future Work}
\label{sec:Conclusions}

In this work, we investigated the occurrence of anti-patterns in microservice-based architectures developed by students in a project-based learning context. Our study analyzed an architecture developed with the contributions of 216 students, identifying both intra-service and inter-service design issues, as well as service interaction problems and other systemic anti-patterns.

The results demonstrate that students are generally able to design cohesive and well-bounded microservices, completing all key functionalities of their applications. Intra-service and inter-service anti-patterns occurred only in limited cases, suggesting a solid understanding of service decomposition principles. At the same time, service interaction and systemic anti-patterns, such as Chatty Services, timeout issues, and missing API Gateway usage, highlight areas where students often prioritize functional implementation over architectural robustness, resilience, and DevOps best practices.

From an educational perspective, these findings provide actionable insights for teaching microservice architecture. Targeted activities such as hands-on exercises on asynchronous communication, design reviews emphasizing inter-service interactions, and practice with DevOps tools can help students better understand the trade-offs between cohesion, scalability, and communication cost. Furthermore, selecting project requirements suitable for microservice decomposition is critical, as overly aggressive decomposition, particularly in multi-team settings, may unintentionally introduce anti-patterns.

Overall, the study shows that project-based learning can effectively engage students in the development and evolution of realistic microservice architectures while highlighting the importance of integrating architectural quality assessment into the curriculum. While some gaps in background and DevOps experience cannot be completely addressed in a single course, systematic exposure across the curriculum is likely to help students progressively acquire the skills necessary to design robust, maintainable, and resilient microservice systems. 
In this same perspective, placing stronger and more explicit emphasis on architectural anti-patterns within a Software Architecture Design course represents a viable solution to strengthen students’ awareness of these topics.

Future work to be addressed includes:  
\begin{itemize}
    
\item  Longitudinal studies: Analyze how students’ understanding of microservice principles and avoidance of anti-patterns evolves across multiple courses or over successive iterations of project-based learning.

\item Tool Support for Feedback: Develop or integrate automated tools to detect anti-patterns in student projects, providing immediate feedback on service decomposition, interaction patterns, and code quality.

\item Broader Dataset: Expand the study to include more students, different institutions, or more diverse project types to validate the generalizability of the findings.

\item Impact of Pedagogical Interventions: Experiment with targeted teaching interventions, such as dedicated labs on asynchronous communication, API Gateway usage, or fault-tolerance patterns, and measure their effect on reducing anti-pattern occurrences.

\item Integration of DevOps Practices: Explore ways to better integrate DevOps practices in the curriculum, including CI/CD pipelines, automated testing, monitoring, and infrastructure-as-code, to strengthen students’ resilience and system reliability skills.

\item Comparison with Professional Systems: Compare anti-pattern occurrences and architectural quality in student projects with real-world microservice systems to identify gaps and best practices that can be taught.

\end{itemize}

\bibliographystyle{ACM-Reference-Format}
\bibliography{bibliography}

@book{Taylor2010,
  author    = {Richard N. Taylor and Nenad Medvidović and Eric M. Dashofy},
  title     = {Software Architecture: Foundations, Theory, and Practice},
  publisher = {Wiley},
  year      = {2010},
  edition   = {1st},
  isbn      = {978-0470167748},
  pages     = {736},
  address   = {Hoboken, NJ, USA},
  url       = {https://www.wiley.com/en-us/Software+Architecture%3A+Foundations%2C+Theory%2C+and+Practice-p-9780470167748},
  note      = {Includes extensive examples, case studies, and discussions on architectural styles, connectors, patterns, and evaluation techniques.}
}

@book{Richardson2018,
  author    = {Richardson, Chris},
  title     = {Microservices Patterns: With examples in Java},
  publisher = {Manning Publications},
  year      = {2018},
  isbn      = {9781617294549},
  pages     = {520},
  note      = {1st edition}
}

@book{Ambler2012,
  author    = {Ambler, Scott W. and Lines, Mark},
  title     = {Disciplined Agile Delivery: A Practitioner's Guide to Agile Software Delivery in the Enterprise},
  publisher = {IBM Press},
  year      = {2012},
  isbn      = {9780132810135},
  pages     = {544},
  note      = {1\textsuperscript{st} edition}
}

@misc{ISO42010,
  author       = {{ISO/IEC/IEEE}},
  title        = {{ISO/IEC/IEEE 42010:2011 Systems and Software Engineering --- Architecture Description}},
  howpublished = {International Standard},
  year         = {2011},
  url          = {http://www.iso-architecture.org/42010/},
  note         = {ISO/IEC JTC1/SC7},
}

@INPROCEEDINGS{Avritzer2025,
  author={Avritzer, Alberto and Janes, Andrea and Trubiani, Catia and Rodrigues, Helena and Cai, Yuanfang and Menasché, Daniel Sadoc and José Abreu de Oliveira, Álvaro},
  booktitle={2025 IEEE 22nd International Conference on Software Architecture (ICSA)}, 
  title={Architecture and Performance Anti-patterns Correlation in Microservice Architectures}, 
  year={2025},
  volume={},
  number={},
  pages={60-71},
  doi={10.1109/ICSA65012.2025.00016}}

@ARTICLE{Parker2023,
  author={Parker, Garrett and Kim, Samuel and Maruf, Abdullah Al and Cerny, Tomas and Frajtak, Karel and Tisnovsky, Pavel and Taibi, Davide},
  journal={IEEE Access}, 
  title={Visualizing Anti-Patterns in Microservices at Runtime: A Systematic Mapping Study}, 
  year={2023},
  volume={11},
  number={},
  pages={4434-4442},
  doi={10.1109/ACCESS.2023.3236165}}

@INPROCEEDINGS{Bakhtin2025,
  author={Bakhtin, Alexander and Esposito, Matteo and Lenarduzzi, Valentina and Taibi, Davide},
  booktitle={2025 IEEE 22nd International Conference on Software Architecture (ICSA)}, 
  title={Network Centrality as a New Perspective on Microservice Architecture}, 
  year={2025},
  volume={},
  number={},
  pages={72-83},
  doi={10.1109/ICSA65012.2025.00017}}

@article{Tighilt2023,
author = {Tighilt, Rafik and Abdellatif, Manel and Trabelsi, Imen and Madern, Lo\"{\i}c and Moha, Naouel and Gu\'{e}h\'{e}neuc, Yann-Ga\"{e}l},
title = {On the maintenance support for microservice-based systems through the specification and the detection of microservice antipatterns},
year = {2023},
issue_date = {Oct 2023},
publisher = {Elsevier Science Inc.},
address = {USA},
volume = {204},
number = {C},
issn = {0164-1212},
url = {https://doi.org/10.1016/j.jss.2023.111755},
doi = {10.1016/j.jss.2023.111755},
journal = {J. Syst. Softw.},
month = oct,
numpages = {16}
}

@inproceedings{Tighilt2020,
author = {Tighilt, Rafik and Abdellatif, Manel and Moha, Naouel and Mili, Hafedh and Boussaidi, Ghizlane El and Privat, Jean and Gu\'{e}h\'{e}neuc, Yann-Ga\"{e}l},
title = {On the Study of Microservices Antipatterns: a Catalog Proposal},
year = {2020},
isbn = {9781450377690},
publisher = {Association for Computing Machinery},
doi = {10.1145/3424771.3424812},
booktitle = {Proceedings of the European Conference on Pattern Languages of Programs 2020},
articleno = {34},
numpages = {13},
series = {EuroPLoP '20}
}

@INPROCEEDINGS{Schirgi2021,
  author={Schirgi, Thomas and Brenner, Eugen},
  booktitle={2021 IEEE 12th International Conference on Software Engineering and Service Science (ICSESS)}, 
  title={Quality Assurance for Microservice Architectures}, 
  year={2021},
  volume={},
  number={},
  pages={76-80},
  doi={10.1109/ICSESS52187.2021.9522227}}

@inproceedings{AgostiniSB24,
  author       = {Michele Agostini and
                  Jacopo Soldani and
                  Antonio Brogi},
  title        = {Detecting and Resolving Bad Organisational Smells for Microservices},
  booktitle    = {Proceedings of the 19th International Conference on Software Technologies,
                  {ICSOFT} 2024, Dijon, France, July 8-10, 2024},
  pages        = {67--78},
  publisher    = {{SCITEPRESS}},
  year         = {2024},
  doi          = {10.5220/0012851200003753},
}

@inproceedings{ZhouHZHSZ22,
  author       = {Xin Zhou and Huang Huang and He Zhang and Xin Huang and Dong Shao and Chenxing Zhong},
  title        = {A Cross–Company Ethnographic Study on Software Teams for DevOps and Microservices: Organization, Benefits, and Issues},
  booktitle    = {ICSE (SEIP)},
  pages        = {1--10},
  year         = {2022},
  publisher    = {IEEE},
  isbn         = {978-1-6654-9590-5},
  doi          = {10.1109/ICSE-SEIP55303.2022.9794010}
}

@article{Sabir2019,
author = {Sabir, Fatima and Palma, Francis and Rasool, Ghulam and Guéhéneuc, Yann-Gaël and Moha, Naouel},
title = {A systematic literature review on the detection of smells and their evolution in object-oriented and service-oriented systems},
journal = {Software: Practice and Experience},
volume = {49},
number = {1},
pages = {3-39},
doi = {https://doi.org/10.1002/spe.2639},
year = {2019}
}

@article{GUO2020,
title = {A review of project-based learning in higher education: Student outcomes and measures},
journal = {International Journal of Educational Research},
volume = {102},
pages = {101586},
year = {2020},
issn = {0883-0355},
doi = {https://doi.org/10.1016/j.ijer.2020.101586},
author = {Pengyue Guo and Nadira Saab and Lysanne S. Post and Wilfried Admiraal},
}

@inproceedings{Fioravanti2018, 
author = {Fioravanti, Maria Lydia and Sena, Bruno and Paschoal, Leo Natan and Silva, La\'{\i}za R. and Allian, Ana P. and Nakagawa, Elisa Y. and Souza, Simone R.S. and Isotani, Seiji and Barbosa, Ellen F.},
title = {Integrating Project Based Learning and Project Management for Software Engineering Teaching: An Experience Report},
year = {2018},
isbn = {9781450351034},
publisher = {Association for Computing Machinery},
doi = {10.1145/3159450.3159599},
booktitle = {Proceedings of the 49th ACM Technical Symposium on Computer Science Education},
pages = {806–811},
numpages = {6},
series = {SIGCSE '18}
}

@InProceedings{Hamer2021,
author="Hamer, Sivana
and Quesada-L{\'o}pez, Christian
and Mart{\'i}nez, Alexandra
and Jenkins, Marcelo",
title="Measuring Students' Source Code Quality in Software Development Projects Through Commit-Impact Analysis",
booktitle="Information Technology and Systems",
year="2021",
publisher="Springer International Publishing",
pages="100--109",
isbn="978-3-030-68418-1",
doi="https://doi.org/10.1007/978-3-030-68418-1_11"
}

@article{Keuning2018,
author = {Keuning, Hieke and Jeuring, Johan and Heeren, Bastiaan},
title = {A Systematic Literature Review of Automated Feedback Generation for Programming Exercises},
year = {2018},
issue_date = {March 2019},
publisher = {Association for Computing Machinery},
volume = {19},
number = {1},
doi = {10.1145/3231711},
journal = {ACM Trans. Comput. Educ.},
month = sep,
articleno = {3},
numpages = {43}
}

@article{Chen2022,
title = {Code-quality evaluation scheme for assessment of student contributions to programming projects},
journal = {Journal of Systems and Software},
volume = {188},
pages = {111273},
year = {2022},
issn = {0164-1212},
doi = {https://doi.org/10.1016/j.jss.2022.111273},
author = {Hsi-Min Chen and Bao-An Nguyen and Chyi-Ren Dow},
}

@InProceedings{DiMeglio2025,
author = {Di Meglio, Sergio and Pontillo, Valeria and Starace, Luigi Libero Lucio},
title = {REST in Pieces: RESTful Design Rule Violations in Student-Built Web Apps},
year = {2025},
isbn = {978-3-032-04206-4},
publisher = {Springer-Verlag},
doi = {10.1007/978-3-032-04207-1_13},

booktitle = {Software Engineering and Advanced Applications: 51st Euromicro Conference, SEAA 2025},
pages = {191–200},
numpages = {10},
}

@book{Clements2002,
author = {Clements, Paul and Garlan, David and Bass, Len and Stafford, Judith and Nord, Robert and Ivers, James and Little, Reed},
title = {Documenting Software Architectures: Views and Beyond},
year = {2002},
isbn = {0201703726},
publisher = {Pearson Education},

}

@book{newman2021building,
  title     = {Building Microservices: Designing Fine-Grained Systems},
  author    = {Newman, Sam},
  year      = {2021},
  publisher = {O’Reilly Media},
  edition   = {2nd}
}

@article{fowler2014microservices,
  title={Microservices},
  author={Fowler, Martin and Lewis, James},
  year={2014},
  journal={martinfowler.com},
  url={https://martinfowler.com/articles/microservices.html}
}

@book{Taylor2009,
author = {Taylor, R. N. and Medvidovic, N. and Dashofy, E. M.},
title = {Software Architecture: Foundations, Theory, and Practice},
year = {2009},
isbn = {0470167742},
publisher = {Wiley Publishing},
}

@inproceedings{Keuning2017,
author = {Keuning, Hieke and Heeren, Bastiaan and Jeuring, Johan},
title = {Code Quality Issues in Student Programs},
year = {2017},
isbn = {9781450347044},
publisher = {Association for Computing Machinery},
doi = {10.1145/3059009.3059061},
booktitle = {Proceedings of the 2017 ACM Conference on Innovation and Technology in Computer Science Education},
pages = {110–115},
numpages = {6},
series = {ITiCSE '17}
}

@inproceedings{Deluca2024,
author = {De Luca, Marco and Di Meglio, Sergio and Fasolino, Anna Rita and Starace, Luigi Libero Lucio and Tramontana, Porfirio},
title = {Automatic Assessment of Architectural Anti-patterns and Code Smells in Student Software Projects},
year = {2024},
isbn = {9798400717017},
publisher = {Association for Computing Machinery},
doi = {10.1145/3661167.3661290},
booktitle = {Proceedings of the 28th International Conference on Evaluation and Assessment in Software Engineering},
pages = {565–569},
numpages = {5},
series = {EASE '24}
}

@InProceedings{DeLuca2025,
author="De Luca, Marco
and Di Martino, Sergio
and Di Meglio, Sergio
and Fasolino, Anna Rita
and Starace, Luigi Libero Lucio
and Tramontana, Porfirio",
title="Rookie Mistakes: Measuring Software Quality in Student Projects to Guide Educational Enhancement",
booktitle="Software Engineering and Advanced Applications",
year="2026",
publisher="Springer Nature Switzerland",
pages="137--154",
isbn="978-3-032-04207-1",
doi={https://doi.org/10.1007/978-3-032-04207-1_10}
}

@book{newman2015building,
  title={Building Microservices},
  author={Newman, Sam},
  year={2015},
  publisher={O'Reilly Media, Inc.}
}

@incollection{dragoni2017microservices,
  title={Microservices: yesterday, today, and tomorrow},
  author={Dragoni, Nicola and Giazzi, Andrea and Lafuente, Alberto Lluch and Mazzara, Manuel and Montesi, Fabrizio and Mustafin, Ruslan and Safina, Larisa},
  booktitle={Present and Ulterior Software Engineering},
  pages={195--216},
  year={2017},
  publisher={Springer},
  doi = {https://doi.org/10.1007/978-3-319-67425-4_12}
}

@article{soldani2018pains,
title = {The pains and gains of microservices: A Systematic grey literature review},
journal = {Journal of Systems and Software},
volume = {146},
pages = {215-232},
year = {2018},
issn = {0164-1212},
doi = {https://doi.org/10.1016/j.jss.2018.09.082},
author = {Jacopo Soldani and Damian Andrew Tamburri and Willem-Jan {Van Den Heuvel}}
}

@INPROCEEDINGS{Lago2005,
  author={Lago, Patricia and van Vliet, Hans},
  booktitle={18th Conference on Software Engineering Education and Training (CSEET'05)}, 
  title={Teaching a Course on Software Architecture}, 
  year={2005},
  volume={},
  number={},
  pages={35-42},
  doi={10.1109/CSEET.2005.33}}

@INPROCEEDINGS{Galster2016,
author = {Galster, Matthias and Angelov, Samuil},
title = {What makes teaching software architecture difficult?},
year = {2016},
isbn = {9781450342056},
publisher = {Association for Computing Machinery},
doi = {10.1145/2889160.2889187},
booktitle = {Proceedings of the 38th International Conference on Software Engineering Companion},
pages = {356–359},
numpages = {4},
series = {ICSE '16}
}

@inbook{Bass2003,
author = {Bass, Len and Clements, Paul and Kazman, Rick},
title = {Software Architecture in Practice},
year = {2012},
isbn = {0321815734},
publisher = {Addison-Wesley Professional},
edition = {3rd}
}

@inproceedings{Taibi2019MicroservicesAntiPatterns,
author="Taibi, Davide
and Lenarduzzi, Valentina
and Pahl, Claus",
title="Microservices Anti-patterns: A Taxonomy",
bookTitle="Microservices: Science and Engineering",
year="2020",
publisher="Springer International Publishing",
pages="111--128",
doi="10.1007/978-3-030-31646-4_5",
}

@article{Cerny2023MicroserviceAntiPatterns,
title = {Catalog and detection techniques of microservice anti-patterns and bad smells: A tertiary study},
journal = {Journal of Systems and Software},
volume = {206},
pages = {111829},
year = {2023},
issn = {0164-1212},
doi = {https://doi.org/10.1016/j.jss.2023.111829},
author = {Tomas Cerny and Amr S. Abdelfattah and Abdullah Al Maruf and Andrea Janes and Davide Taibi}
}

@ARTICLE{8354414,
  author={Taibi, Davide and Lenarduzzi, Valentina},
  journal={IEEE Software}, 
  title={On the Definition of Microservice Bad Smells}, 
  year={2018},
  volume={35},
  number={3},
  pages={56-62},
  doi={10.1109/MS.2018.2141031}}

@book{fowler1999refactoring,
  title={Refactoring: Improving the design of existing code},
  author={Fowler, Martin and Beck, Kent and Brant, John and Opdyke, William and Roberts, Erich},
  year={1999},
  publisher={Addison-Wesley}
}

@INPROCEEDINGS{CodeOwnershipPrinciples2024,
  author={Thongtanunam, Patanamon and Tantithamthavorn, Chakkrit},
  booktitle={2024 IEEE 35th International Symposium on Software Reliability Engineering (ISSRE)}, 
  title={Code Ownership: The Principles, Differences, and Their Associations with Software Quality}, 
  year={2024},
  volume={},
  number={},
  pages={379-390},
  doi={10.1109/ISSRE62328.2024.00044}}

@inproceedings{CodeOwnershipAISecurity2023,
author = {Wen, Jiawen and Yuan, Dong and Ma, Lei and Chen, Huaming},
title = {Code Ownership in Open-Source AI Software Security},
year = {2024},
isbn = {9798400705724},
publisher = {Association for Computing Machinery},
url = {https://doi.org/10.1145/3643691.3648586},
booktitle = {Proceedings of the 2nd International Workshop on Responsible AI Engineering},
pages = {28–35},
numpages = {8},
series = {RAIE '24}
}

@inproceedings{AlignmentOwnershipContribution2019,
  author={Zabardast, Ehsan and Gonzalez-Huerta, Javier and Tanveer, Binish},
  booktitle={2022 IEEE 19th International Conference on Software Architecture Companion (ICSA-C)}, 
  title={Ownership vs Contribution: Investigating the Alignment Between Ownership and Contribution}, 
  year={2022},
  volume={},
  number={},
  pages={30-34},
  doi={10.1109/ICSA-C54293.2022.00013}}

@misc{SourcegraphVision2023,
  author       = {Quinn Trotter and Erik Kesty and Quinn Bays},
  title        = {Our vision for better code ownership},
  year         = {2023},
  day          = {13},
  note         = {Sourcegraph Blog},
  url          = {https://sourcegraph.com/blog/our-vision-for-code-ownership}
}

@article{BALDASSARRE2020106377,
title = {On the diffuseness of technical debt items and accuracy of remediation time when using SonarQube},
journal = {Information and Software Technology},
volume = {128},
pages = {106377},
year = {2020},
issn = {0950-5849},
doi = {https://doi.org/10.1016/j.infsof.2020.106377},
author = {Maria Teresa Baldassarre and Valentina Lenarduzzi and Simone Romano and Nyyti Saarimäki}
}

@article{ALFAYEZ2023107147,
title = {How SonarQube-identified technical debt is prioritized: An exploratory case study},
journal = {Information and Software Technology},
volume = {156},
pages = {107147},
year = {2023},
issn = {0950-5849},
doi = {https://doi.org/10.1016/j.infsof.2023.107147},
author = {Reem Alfayez and Robert Winn and Wesam Alwehaibi and Elaine Venson and Barry Boehm}
}

@article{Ferreira2025TeachingMS,
  title     = {Teaching Complex Systems Based on Microservices},
      author={Renato Cordeiro Ferreira and Thatiane de Oliveira Rosa and Alfredo Goldman and Eduardo Guerra},
  journal   = {arXiv preprint},
  year      = {2025},
  url       = {https://arxiv.org/abs/2506.16492}
}

@inproceedings{TeachingMSADevOps2022,
author = {B\ae{}rbak Christensen, Henrik},
title = {Teaching Microservice Architecture Using DevOps—An Experience Report},
year = {2022},
isbn = {978-3-031-16696-9},
publisher = {Springer-Verlag},
doi = {10.1007/978-3-031-16697-6_8},
booktitle = {Software Architecture: 16th European Conference, ECSA 2022},
pages = {117–130},
numpages = {14},
}

@article{Qian2025OBEMicroservices,
  title     = {Research on Teaching Reform of Spring Cloud Microservice Architecture Based on OBE Concept},
  author    = {Zihao, Qian},
  journal   = {Advances in Educational Technology and Psychology},
  year      = {2025},
  volume    = {9},
  number    = {2},
  publisher = {Clausius Press},
  url       = {https://dx.doi.org/10.23977/curtm.2025.080420}
}

@inproceedings{Meissner2021EAsLiT,
  title     = {Flexible educational software architecture: at the example of EAs.LiT 2},
  booktitle = {International Workshop on Intelligent Mentoring in Higher Education (IMHE)},
  author    = {Meissner, Roy and Thor, Andreas},
  year      = {2020},
  url       = {https://ceur-ws.org/Vol-3046/imhe_2020_paper_2.pdf}
}

@INPROCEEDINGS{maruf2022telemetry,
  author={Al Maruf, Abdullah and Bakhtin, Alexander and Cerny, Tomas and Taibi, Davide},
  booktitle={2022 IEEE International Conference on Service-Oriented System Engineering (SOSE)}, 
  title={Using Microservice Telemetry Data for System Dynamic Analysis}, 
  year={2022},
  volume={},
  number={},
  pages={29-38},
  doi={10.1109/SOSE55356.2022.00010}}

@inproceedings{niedermaier2019observability,
author = {Niedermaier, Sina and Koetter, Falko and Freymann, Andreas and Wagner, Stefan},
title = {On Observability and Monitoring of Distributed Systems – An Industry Interview Study},
year = {2019},
isbn = {978-3-030-33701-8},
publisher = {Springer-Verlag},
doi = {10.1007/978-3-030-33702-5_3},
booktitle = {Service-Oriented Computing: 17th International Conference, ICSOC 2019},
pages = {36–52},
numpages = {17},
}

@book{brown1998antipatterns,
  title={AntiPatterns: Refactoring Software, Architectures, and Projects in Crisis},
  author={Brown, William J. and Malveau, Raphael C. and McCormick III, Hays W. and Mowbray, Thomas J.},
  year={1998},
  publisher={Wiley}
}

@article{Offline_Mining,
author = {Soldani, Jacopo and Khalili, Javad and Brogi, Antonio},
year = {2023},
month = {04},
pages = {},
title = {Offline Mining of Microservice-Based Architectures (Extended Version)},
volume = {4},
journal = {SN Computer Science},
doi = {10.1007/s42979-023-01721-4}
}

@book{wohlin2024experimentation,
  title        = {Experimentation in Software Engineering},
  author       = {Claes Wohlin and Per Runeson and Martin Höst and Magnus C. Ohlsson and Björn Regnell and Anders Wesslén},
  year         = {2024},
  edition      = {2},
  publisher    = {Springer Berlin, Heidelberg},
  doi          = {10.1007/978-3-662-69306-3},
  isbn         = {978-3-662-69306-3},
  isbn2        = {978-3-662-69305-6},
  pages        = {XXV, 274}
}

@article{Cerny_tool,
  author  = {Simon Schneider and Alexander Bakhtin and Xiaozhou Li and Jacopo Soldani and Antonio Brogi and Tomas Cerny and Riccardo Scandariato and Davide Taibi},
  title   = {Comparison of static analysis architecture recovery tools for microservice applications},
  journal = {Empirical Software Engineering},
  year    = {2025},
  volume  = {30},
  number  = {5},
  pages   = {128},
  doi     = {10.1007/s10664-025-10686-2},
  issn    = {1573-7616},
  month   = {jun}
}

@Article{Bushong21,
AUTHOR = {Bushong, Vincent and Abdelfattah, Amr S. and Maruf, Abdullah A. and Das, Dipta and Lehman, Austin and Jaroszewski, Eric and Coffey, Michael and Cerny, Tomas and Frajtak, Karel and Tisnovsky, Pavel and Bures, Miroslav},
TITLE = {On Microservice Analysis and Architecture Evolution: A Systematic Mapping Study},
JOURNAL = {Applied Sciences},
VOLUME = {11},
YEAR = {2021},
NUMBER = {17},
ARTICLE-NUMBER = {7856},
ISSN = {2076-3417},
DOI = {10.3390/app11177856}
}

@article{BRANDON20,
title = {Graph-based root cause analysis for service-oriented and microservice architectures},
journal = {Journal of Systems and Software},
volume = {159},
pages = {110432},
year = {2020},
issn = {0164-1212},
doi = {https://doi.org/10.1016/j.jss.2019.110432},
author = {Alvaro Brandón and Marc Solé and Alberto Huélamo and David Solans and María S. Pérez and Victor Muntés-Mulero}
}

@inproceedings{Liu19,
author = {Liu, Haifeng and Zhang, Jinjun and Shan, Huasong and Li, Min and Chen, Yuan and He, Xiaofeng and Li, Xiaowei},
title = {JCallGraph: Tracing Microservices in Very Large Scale Container Cloud Platforms},
year = {2019},
isbn = {978-3-030-23501-7},
publisher = {Springer-Verlag},
doi = {10.1007/978-3-030-23502-4_20},
booktitle = {Cloud Computing – CLOUD 2019: 12th International Conference},
pages = {287–302},
numpages = {16},
}

@INPROCEEDINGS{Mendonca20,
  author={Mendonca, Nabor C. and Aderaldo, Carlos M. and Camara, Javier and Garlan, David},
  booktitle={2020 IEEE International Conference on Software Architecture (ICSA)}, 
  title={Model-Based Analysis of Microservice Resiliency Patterns}, 
  year={2020},
  volume={},
  number={},
  pages={114-124},
  doi={10.1109/ICSA47634.2020.00019}}

@INPROCEEDINGS{McZara20,
  author={McZara, Jason and Kafle, Subash and Shin, Daniel},
  booktitle={2020 IEEE International Conference on Smart Cloud (SmartCloud)}, 
  title={Modeling and Analysis of Dependencies between Microservices in DevSecOps}, 
  year={2020},
  volume={},
  number={},
  pages={140-147},
  doi={10.1109/SmartCloud49737.2020.00034}
}

@article{Walker20,
AUTHOR = {Walker, Andrew and Das, Dipta and Cerny, Tomas},
TITLE = {Automated Code-Smell Detection in Microservices Through Static Analysis: A Case Study},
JOURNAL = {Applied Sciences},
VOLUME = {10},
YEAR = {2020},
NUMBER = {21},
ARTICLE-NUMBER = {7800},
ISSN = {2076-3417},
DOI = {10.3390/app10217800}
}

@inproceedings{Marquez19,
author = {Marquez, Gaston and Astudillo, Hernan},
title = {Identifying availability tactics to support security architectural design of microservice-based systems},
year = {2019},
isbn = {9781450371421},
publisher = {Association for Computing Machinery},
doi = {10.1145/3344948.3344996},
booktitle = {Proceedings of the 13th European Conference on Software Architecture - Volume 2},
pages = {123–129},
numpages = {7},
series = {ECSA '19}
}

@inproceedings{Pigazzini2020,
author = {Pigazzini, Ilaria and Fontana, Francesca Arcelli and Lenarduzzi, Valentina and Taibi, Davide},
title = {Towards microservice smells detection},
year = {2020},
isbn = {9781450379601},
publisher = {Association for Computing Machinery},
doi = {10.1145/3387906.3388625},
booktitle = {Proceedings of the 3rd International Conference on Technical Debt},
pages = {92–97},
numpages = {6},
series = {TechDebt '20}
}

@inproceedings{Lau2024,
author = {Lau, Yi Meng and Koh, Christian Michael and Jiang, Lingxiao},
title = {Teaching Software Development for Real-World Problems using a Microservice-Based Collaborative Problem-Solving Approach},
year = {2024},
isbn = {9798400704987},
publisher = {Association for Computing Machinery},
doi = {10.1145/3639474.3640064},
booktitle = {Proceedings of the 46th International Conference on Software Engineering: Software Engineering Education and Training},
pages = {22–33},
numpages = {12},
series = {ICSE-SEET '24}
}

@inproceedings{Meissner2025,
author = {Mei\ss{}ner, Niklas and Bredl, Paul and Speth, Sandro and Becker, Steffen},
title = {Enhancing Motivation in Software Engineering Education through Gamified Agile Project-based Learning},
year = {2025},
isbn = {9798400712760},
publisher = {Association for Computing Machinery},
doi = {10.1145/3696630.3727241},
booktitle = {Proceedings of the 33rd ACM International Conference on the Foundations of Software Engineering},
pages = {835–846},
numpages = {12},
series = {FSE Companion '25}
}

@inproceedings{Souza19,
author = {Souza, Maur\'{\i}cio and Moreira, Renata and Figueiredo, Eduardo},
title = {Students Perception on the use of Project-Based Learning in Software Engineering Education},
year = {2019},
isbn = {9781450376518},
publisher = {Association for Computing Machinery},
doi = {10.1145/3350768.3352457},
booktitle = {Proceedings of the XXXIII Brazilian Symposium on Software Engineering},
pages = {537–546},
numpages = {10},
series = {SBES '19}
}

@inproceedings{perez20,
author = {P\'{e}rez, Beatriz and Rubio, \'{A}ngel L.},
title = {A Project-Based Learning Approach for Enhancing Learning Skills and Motivation in Software Engineering},
year = {2020},
isbn = {9781450367936},
publisher = {Association for Computing Machinery},
doi = {10.1145/3328778.3366891},
booktitle = {Proceedings of the 51st ACM Technical Symposium on Computer Science Education},
pages = {309–315},
numpages = {7},
series = {SIGCSE '20}
}

@INPROCEEDINGS{Arcan_tool,
  author={Bacchiega, Paolo and Rusconi, Davide and Mereghetti, Paolo and Fontana, Francesca Arcelli},
  booktitle={2024 IEEE 21st International Conference on Software Architecture Companion (ICSA-C)}, 
  title={Refactoring of a Microservices Project Driven by Architectural Smell Detection}, 
  year={2024},
  volume={},
  number={},
  pages={281-288},
  doi={10.1109/ICSA-C63560.2024.00040}}

@article{ci_bad_practices,
author = {Zampetti, Fiorella and Vassallo, Carmine and Panichella, Sebastiano and Canfora, Gerardo and Gall, Harald and Di Penta, Massimiliano},
year = {2020},
month = {03},
pages = {},
title = {An Empirical Characterization of Bad Practices in Continuous Integration},
volume = {25},
journal = {Empirical Software Engineering},
doi = {10.1007/s10664-019-09785-8}
}

@article{ponce2022smells,
    author  = "Fabián Ponce and Jacopo Soldani and Hugo Astudillo and Antonio Brogi",
    title   = "Smells and Refactorings for Microservices Security: A Multivocal Literature Review",
    journal = "Journal of Systems and Software",
    volume  = "192",
    pages   = "111393",
    year    = "2022",
    issn    = "0164-1212",
    doi     = "10.1016/j.jss.2022.111393",
    url     = "https://www.sciencedirect.com/science/article/pii/S016412122200111X"
}

\end{document}